\documentclass[trackchanges, twocolumn]{aastex701}

\usepackage{booktabs}

\newcommand{\um}{\,$\mu$m}
\newcommand{\kms}{\,km\,s$^{-1}$}
\newcommand{\snia}{SN\,Ia}
\newcommand{\rbs}{SN\,2025rbs}

\newcommand{\CIERA}{\affiliation{Center for Interdisciplinary Exploration and Research in Astrophysics (CIERA), Northwestern University, Evanston, IL 60201, USA}}

\newcommand{\Konkoly}{\affiliation{Konkoly Observatory, HUN\textendash REN Research Center for Astronomy and Earth Sciences, Budapest, 1121, Hungary}}
\newcommand{\KyotoU}{\affiliation{Department of Astronomy, Kyoto University, Kyoto 606\textendash 8502, Japan}}
\newcommand{\LCO}{\affiliation{Las Cumbres Observatory, Goleta, CA 93117, USA}}
\newcommand{\Northwestern}{\affiliation{Department of Physics and Astronomy, Northwestern University, Evanston, IL 60208, USA}}
\newcommand{\Princeton}{\affiliation{Department of Astrophysical Sciences, Princeton University, Princeton, NJ 08544, USA}}

\newcommand{\Rutgers}{\affiliation{Department of Physics and Astronomy, Rutgers, The State University of New Jersey, Piscataway, NJ 08854, USA}}
\newcommand{\SkAI}{\affiliation{NSF\textendash Simons AI Institute for the Sky (SkAI), Chicago, IL 60611, USA}}

\newcommand{\UA}{\affiliation{Steward Observatory, University of Arizona, Tucson, AZ 85721, USA}}
\newcommand{\UCD}{\affiliation{Department of Physics and Astronomy, University of California, Davis, CA 95616, USA}}
\newcommand{\UCSC}{\affiliation{Department of Astronomy and Astrophysics, University of California, Santa Cruz, CA 95064, USA}}

\newcommand{\USzeged}{\affiliation{Department of Experimental Physics, Institute of Physics, University of Szeged, D\'om t\'er 9, Szeged, 6720 Hungary}}
\newcommand{\UTexas}{\affiliation{Department of Astronomy, The University of Texas at Austin, Austin, TX 78712, USA}}
\newcommand{\ESOgermany}{\affiliation{European Southern Observatory, Karl-Schwarzschild-Stra\ss{}e 2, D-85748, Garching bei M\"unchen, Germany}}
\newcommand{\AixMarseille}{\affiliation{Aix Marseille Univ, CNRS, CNES, LAM, Marseille, France}}
\newcommand{\UCSB}{\affiliation{Department of Physics, University of California, Santa Barbara, CA 93106-9530, USA}}
\newcommand{\TUM}{\affiliation{Technical University of Munich, TUM School of Natural Sciences, Physics Department, D-85748, Garching bei M\"unchen, Germany}}

\newcommand{\Hawaii}{\affiliation{Institute for Astronomy, University of Hawai'i at M\=anoa, 2680 Woodlawn Dr., Hawai'i, HI 96822, USA}}
\newcommand{\UCBerkeley}{\affiliation{Department of Astronomy, University of California, Berkeley, CA 94720-3411, USA}}
\newcommand{\MillerInstitute}{\affiliation{Miller Institute for Basic Research in Science, 206B Stanley Hall, Berkeley, CA 94720, USA}}
\newcommand{\Portsmouth}{\affiliation{Institute of Cosmology and Gravitation, University of Portsmouth, Portsmouth, PO1 3FX, UK}}
\newcommand{\AMNH}{\affiliation{Department of Astrophysics, American Museum of Natural History, New York, NY 10024, USA}}
\newcommand{\TCD}{\affiliation{School of Physics, Trinity College Dublin, The University of Dublin, Dublin 2, Ireland}}
\newcommand{\UIUC}{\affiliation{Department of Astronomy, University of Illinois Urbana-Champaign, 1002 West Green Street, Urbana, IL 61801, USA}}
\newcommand{\MPA}{\affiliation{Max-Planck-Institut f\"ur Astrophysik, Karl-Schwarzschild-Str. 1, D-85748, Garching, Germany}}
\newcommand{\Kanazawa}{\affiliation{College of Science and Engineering, Kanazawa University, Kanazawa, Japan}}
\newcommand{\Monash}{\affiliation{School of Physics and Astronomy, Monash University, Clayton, Victoria 3800, Australia}}
\newcommand{\OzGrav}{\affiliation{OzGrav: The ARC Centre of Excellence for Gravitational Wave Discovery, Clayton, Victoria 3800, Australia}}
\newcommand{\KyotoOkayama}{\affiliation{Okayama Observatory, Astronomical Observatory, Graduate School of Science, Kyoto University, 3037-5 Honjo, Kamogata-cho, Asakuchi City, Okayama 719-0232, Japan}}
\newcommand{\QUB}{\affiliation{School of Mathematics and Physics, Queen's University Belfast, University Road, Belfast BT7 1NN, UK}}
\newcommand{\Melbourne}{\affiliation{School of Physics, The University of Melbourne, VIC 3010, Australia}}
\newcommand{\IIA}{\affiliation{Indian Institute of Astrophysics, II Block, Koramangala, Bangalore, India}}
\newcommand{\GeminiObs}{\affiliation{Gemini Observatory/NSF's NOIRLab, 670 N. A'ohoku Place, Hilo, HI 96720, USA}}
\newcommand{\Tsinghua}{\affiliation{Physics Department, Tsinghua University, Beijing, 100084, China}}
\newcommand{\GSI}{\affiliation{GSI Helmholtzzentrum f\"ur Schwerionenforschung GmbH, 64291 Darmstadt, Germany}}
\newcommand{\STScI}{\affiliation{Space Telescope Science Institute, 3700 San Martin Drive, Baltimore, MD 21218, USA}}
\newcommand{\JHU}{\affiliation{Department of Physics and Astronomy, The Johns Hopkins University, Baltimore, MD 21218, USA}}
\newcommand{\HawaiiHilo}{\affiliation{Institute for Astronomy, University of Hawai'i, 640 N. A'ohoku Pl., Hilo, HI 96720, USA}}

\begin{document}

\title{JWST Spectroscopy of Type\,Ia Supernova 2025rbs from Maximum Light to the Nebular Phase}

\author[orcid=0000-0003-3108-1328,sname='Kwok']{Lindsey~A.~Kwok}
\thanks{NHFP Hubble Fellow}
\CIERA
\email[show]{lindsey.kwok@northwestern.edu}

\author[0000-0002-9388-2932]{St\'{e}phane Blondin}
\ESOgermany
\AixMarseille
\email{stephane.blondin@eso.org}

\author[0000-0001-9515-478X]{Adam~A.~Miller}
\Northwestern
\CIERA
\SkAI
\email{amiller@northwestern.edu}

\author[0000-0001-8738-6011]{Saurabh W.\ Jha}
\Rutgers
\email{saurabh@physics.rutgers.edu}

\author[orcid=0000-0003-3953-9532,sname='Hoogendam']{Willem~B.~Hoogendam}
\Hawaii
\email{willemh@hawaii.edu}

\author[orcid=0000-0002-7305-8321,sname='Pfeffer']{Cameron~M.~Pfeffer}
\Hawaii
\email{cpfeffer@hawaii.edu}


\author[orcid=0009-0000-7818-9817,sname='Abate']{Eyouel~Z.~Abate}
\UCBerkeley
\email{eyouelabate@berkeley.edu}

\author[orcid=0000-0003-0123-0062,sname='Andrews']{Jennifer~E.~Andrews}
\GeminiObs
\email{jennifer.andrews@noirlab.edu}

\author[orcid=0000-0002-1895-6639,sname='Andrews']{Moira~Andrews}
\LCO
\UCSB
\email{mandrews@lco.global}

\author[orcid=0000-0002-5221-7557,sname='Ashall']{Chris~Ashall}
\Hawaii
\email{cashall@hawaii.edu}

\author[orcid=0000-0002-4449-9152,sname='Auchettl']{Katie~Auchettl}
\Melbourne
\email{katie.auchettl@unimelb.edu.au}

\author[orcid=0009-0007-1739-1414,sname='Betz']{Andreas~Betz}
\UCBerkeley
\email{andreasbetz@berkeley.edu}

\author[orcid=0000-0002-4924-444X,sname='Bostroem']{K.~Azalee~Bostroem}
\thanks{LSST-DA Catalyst Fellow}
\UA
\email{bostroem@arizona.edu}

\author[orcid=0000-0001-5955-2502,sname='Brink']{Thomas~G.~Brink}
\UCBerkeley
\email{tgbrink@berkeley.edu}

\author[orcid=0000-0002-7975-8185,sname='Callan']{Fionntan~P.~Callan}
\QUB
\email{f.callan@qub.ac.uk}

\author[orcid=0000-0003-0528-202X,sname='Christy']{Collin~T.~Christy}
\UA
\email{collinchristy@arizona.edu}

\author[orcid=0000-0003-4914-5625,sname='Farah']{Joseph~R.~Farah}
\LCO
\UCSB
\MillerInstitute
\UCBerkeley
\email{josephfarah@berkeley.edu}

\author[orcid=0000-0003-3460-0103,sname='Filippenko']{Alexei~V.~Filippenko}
\UCBerkeley
\email{afilippenko@berkeley.edu}

\author[orcid=0000-0003-2024-2819,sname='Flors']{Andreas~Fl{\"o}rs}
\GSI
\email{a.floers@gsi.de}

\author[orcid=0000-0002-2445-5275,sname='Foley']{Ryan~J.~Foley}
\UCSC
\email{foley@ucsc.edu}

\author[orcid=0009-0002-8671-4858,sname='Gendreau-Distler']{Eli~Gendreau-Distler}
\UCBerkeley
\email{egendreaudistler@berkeley.edu}

\author[orcid=0000-0002-4391-6137,sname='Graur']{Or~Graur}
\Portsmouth
\AMNH
\email{or.graur@port.ac.uk}

\author[orcid=0000-0001-9668-2920,sname='Hinkle']{Jason~T.~Hinkle}
\thanks{NHFP Einstein Fellow}
\UIUC
\SkAI
\email{jhinkle6@illinois.edu}

\author[orcid=0000-0003-4253-656X,sname='Howell']{D.~Andrew~Howell}
\LCO
\UCSB
\email{ahowell@lco.global}

\author[orcid=0000-0002-6230-0151,sname='Jones']{David~O.~Jones}
\HawaiiHilo
\email{dojones@hawaii.edu}

\author[orcid=0009-0002-3925-0536,sname='Kageyama']{Rinon~Kageyama}
\Kanazawa
\email{rinon.kageyama@astro.s.kanazawa-u.ac.jp}

\author[orcid=0000-0002-4540-4928,sname='Kawabata']{Miho~Kawabata}
\KyotoOkayama
\email{kawabata@kusastro.kyoto-u.ac.jp}

\author[orcid=0009-0000-9117-8995,sname='Koelln']{Cristine~Koelln}
\ESOgermany
\TUM
\email{cristine.koelln@eso.org}

\author[orcid=0000-0003-2037-4619,sname='Larison']{Conor~Larison}
\Rutgers
\email{conorjlarison@gmail.com}

\author[orcid=0000-0002-7866-4531,sname='Liu']{Chang~Liu}
\Northwestern
\CIERA
\SkAI
\email{ptg.cliu@u.northwestern.edu}

\author[orcid=0000-0003-2611-7269,sname='Maeda']{Keiichi~Maeda}
\KyotoU
\email{keiichi.maeda@kusastro.kyoto-u.ac.jp}

\author[orcid=0000-0002-9770-3508,sname='Maguire']{Kate~Maguire}
\TCD
\email{kate.maguire@tcd.ie}

\author[orcid=0000-0001-5807-7893,sname='McCully']{Curtis~McCully}
\LCO
\email{cmccully@lco.global}

\author[orcid=0000-0001-7186-105X,sname='Medler']{Kyle~Medler}
\Hawaii
\email{kmedler@hawaii.edu}

\author[orcid=0000-0002-7015-3446,sname='Meza-Retamal']{Nicolas~E.~Meza-Retamal}
\UCD
\email{nicomezare@gmail.com}

\author[orcid=0009-0007-7258-3072,sname='Mina']{Ann~Mina}
\UCBerkeley
\email{annvictor2004@berkeley.edu}

\author[orcid=0000-0003-3308-2420,sname='Pakmor']{R\"udiger~Pakmor}
\MPA
\email{rpakmor@mpa-garching.mpg.de}

\author[, sname='Parke']{Ansel Parke}
\UCBerkeley
\email{aparke@berkeley.edu}

\author[orcid=0009-0009-1267-8445,sname='Patlak']{Riley~Patlak}
\UCBerkeley
\email{rileypatlak@berkeley.edu}

\author[orcid=0000-0002-0744-0047,sname='Pearson']{Jeniveve~Pearson}
\UA
\email{jenivevepearson@arizona.edu}

\author[orcid=0000-0002-7352-7845,sname='Ravi']{Aravind~P.~Ravi}
\UCD
\email{apazhayathravi@ucdavis.edu}

\author[orcid=0000-0002-5683-2389,sname='Rehemtulla']{Nabeel~Rehemtulla}
\Northwestern
\CIERA
\SkAI
\email{nabeelrehemtulla2027@u.northwestern.edu}

\author[orcid=0000-0002-4410-5387,sname='Rest']{Armin~Rest}
\STScI
\JHU
\email{arest@stsci.edu}

\author[orcid=0000-0003-4102-380X,sname='Sand']{David~J.~Sand}
\UA
\email{dsand@arizona.edu}

\author[orcid=0000-0001-8023-4912,sname='Sears']{Huei~Sears}
\Rutgers
\email{huei.sears@rutgers.edu}

\author[orcid=0000-0003-4631-1149,sname='Shappee']{Benjamin~J.~Shappee}
\Hawaii
\email{shappee@hawaii.edu}

\author[orcid=0000-0002-4022-1874,sname='Shrestha']{Manisha~Shrestha}
\Monash
\OzGrav
\email{manisha.shrestha@monash.edu}

\author[orcid=0000-0001-6706-2749,sname='Singh']{Mridweeka~Singh}
\IIA
\email{yashasvi04@gmail.com}

\author[orcid=0000-0003-4610-1117,sname='Szalai']{Tam\'as~Szalai}
\USzeged
\email{szaszi@titan.physx.u-szeged.hu}

\author[orcid=0000-0002-8482-8993,sname='Taguchi']{Kenta~Taguchi}
\KyotoOkayama
\email{kentagch@kusastro.kyoto-u.ac.jp}

\author[orcid=0000-0001-7380-3144,sname='Temim']{Tea~Temim}
\Princeton
\email{tea.temim@gmail.com}

\author[orcid=0000-0001-9834-3439,sname='Terwel']{Jacco~H.~Terwel}
\TCD
\email{terwelj@tcd.ie}

\author[orcid=0000-0001-8818-0795,sname='Valenti']{Stefano~Valenti}
\UCD
\email{valenti@ucdavis.edu}

\author[orcid=0000-0001-8764-7832,sname='Vinko']{J\'ozsef~Vink\'o}
\Konkoly
\email{vinko@konkoly.hu}

\author[orcid=0000-0003-1349-6538,sname='Wheeler']{J.~Craig~Wheeler}
\UTexas
\email{wheel@astro.as.utexas.edu}

\author[orcid=0009-0006-7296-728X,sname='Wynn']{Kathryn~Wynn}
\LCO
\UCSB
\email{kwynn@lco.global}

\author[orcid=0000-0002-6535-8500,sname='Yang']{Yi~Yang}
\Tsinghua
\email{yi_yang@mail.tsinghua.edu.cn}

\author[orcid=0000-0002-2636-6508,sname='Zheng']{WeiKang~Zheng}
\UCBerkeley
\email{weikang@berkeley.edu}

\begin{abstract}

We present \textit{JWST} observations of the Type Ia supernova (\snia) 2025rbs ($D=$14.5\,Mpc) at $+$1, $+$23, and $+$84\,days after \textit{B}-band maximum, spanning a wavelength-dependent transition toward the nebular phase. Combined with ground-based optical and near-infrared (NIR) data, our panchromatic spectra (0.4--14\um) include the first maximum-light mid-infrared (MIR) spectrum and the earliest MIR spectroscopic sequence of an \snia\ to date. At peak light, the MIR spectrum exhibits a continuum with permitted and forbidden features, including \ion{Si}{2}, \ion{Ni}{2}, \ion{Mg}{2}, and early-emerging [Ni~{\small III--IV}] and [Ar~{\small II--III}]. By $+$23\,days the MIR is dominated by forbidden lines with a weak continuum, and by $+$84\,days it is fully nebular, whereas the optical/NIR spectra remain transitional. The nebular spectrum reveals strongly stratified ejecta, with stable Ni concentrated at the lowest velocities, radioactive Co at intermediate velocities but absent within $\sim2000$\kms, and Ar occupying an outer shell. We detect small-scale substructure in [\ion{Ca}{4}]\,3.21\um\ with few-percent amplitudes on a characteristic scale of $\sim800$\kms, reflecting clumping in composition or ionization. Radiative-transfer calculations at $+$1\,day substantially underpredict the MIR \ion{Mg}{2} features (4.6, 9.6, and 13.0\um) despite approximately reproducing the NIR \ion{Mg}{2}\,1.0927\um\ line, suggesting that the relative strengths of the NIR and MIR transitions are sensitive to the treatment of Mg ionization and excitation. These observations demonstrate that MIR spectroscopy beginning near maximum light simultaneously probes the emerging inner ejecta and rapidly fading outer burning products, providing new constraints for explosion and radiative-transfer models.

\end{abstract}



\section{Introduction \label{sec:intro}} 

The evolution of Type Ia supernovae (\snia)---thermonuclear explosions of carbon-oxygen white dwarfs \citep[WDs;][]{Hoyle1960}---has been observed extensively in the optical, from the photospheric phase in the weeks surrounding maximum light through the optically thin nebular phase at $\gtrsim$100\,days post peak. However, optical spectra alone provide an incomplete picture: because the opacity is strongly wavelength-dependent \citep{Pinto2000}, different wavelength ranges probe different layers of the ejecta simultaneously. 

At early times, the ejecta layers sampled by a given spectral feature depend on the line opacity as well as the ionization, excitation, and abundance structure. Optically thick permitted lines form above the photosphere and can sample high-velocity material, including the outer ejecta, while wavelength regions with lower effective opacity can provide access to deeper layers. This effect is especially important in the ultraviolet (UV), where heavy line blanketing makes early-time spectra sensitive to the composition and density structure of the outer ejecta \citep[e.g.,][]{Sauer2008, Foley2012, Foley2016, Hachinger2013, DerKacy2020}. At a given phase, near- and mid-infrared (NIR/MIR) spectroscopy can probe different ejecta depths than the optical because of the wavelength-dependent opacity; in many cases, these transitions provide access to deeper layers and burning products that are blended or obscured at shorter wavelengths \citep[e.g.,][]{Wheeler1998, Hoflich2002, Marion2003, Marion2009, Hoogendam25_epr}. Panchromatic spectral coverage enables simultaneous constraints on the ejecta structure at a single epoch. In particular, MIR wavelengths can reveal forbidden emission lines from the interior ejecta even when the optical spectrum still appears photospheric \citep{Kwok2025a}, offering an earlier window into the ejecta interior than optical or NIR observations alone.

Between the photospheric and nebular phases lies a transitional regime in which the ejecta are partially optically thin, permitted and forbidden lines coexist, and the spectrum encodes information from intermediate ejecta depths \citep[e.g.,][]{Friesen2014, Friesen2017}. Substantial optical spectroscopic data exist through this transition, but the phase remains challenging to model because photospheric-like line scattering, continuum opacity, nonthermal excitation/ionization, and forbidden-line cooling can all contribute simultaneously. Time-dependent non-local-thermodynamic-equilibrium (NLTE) radiative-transfer codes can treat broad phase ranges and have modeled optical/NIR spectra through this transition \citep[e.g.,][]{Dessart2014, Blondin2015}, but new panchromatic data provide additional constraints on the relevant physics. MIR observations during this phase reveal species, ionization states, and ejecta depths that are blended, weak, or inaccessible at optical wavelengths, motivating further development and validation of panchromatic radiative-transfer models \citep[e.g.,][]{Blondin2023,Kwok2025a}.

MIR wavelengths are particularly well-suited to studying \snia\ ejecta because they harbor relatively isolated transitions from multiple ionization states of stable and radioactive iron-group elements (IGEs; e.g., Fe, Co, Ni), intermediate-mass elements (IMEs; e.g., S, Si, Ar, Ca), and low-mass elements (LMEs; e.g., Ne, Mg), providing a comprehensive probe of the chemical and ionization structure of the ejecta \citep{Gerardy2007, Kwok2023, Kwok2024, Kwok2025a, Kwok2026, Blondin2023, DerKacy2023, DerKacy2024, Ashall2024, Macrie2026}. At early times, MIR forbidden lines emerge well before their optical counterparts, and at later times the nebular MIR emission lines encode information about the geometry of the emission \citep{Jerkstrand2017}, with implications for the explosion mechanism, that is difficult to disentangle from optical spectra alone. {\it JWST} is transforming the landscape of MIR observations of \snia\ with unprecedented sensitivity and spectral coverage, placing new constraints on ejecta geometry, nucleosynthesis, and explosion mechanisms across a growing sample of events.

Prior to \textit{JWST}, MIR spectroscopy of \snia\ was limited to a handful of observations \citep{Gerardy2007, Telesco2015}. \cite{Telesco2015} obtained the first MIR spectral sequence of an \snia\ during the transitional phase of SN\,2014J, using CanariCam on the Gran Telescopio Canarias (GTC) at phases of $+$37 to $+$117\,days post-maximum (57 to 137\,days post-explosion). These ground-based MIR observations were possible only owing to the exceptional proximity of SN\,2014J ($D\approx3$\,Mpc), and the wavelength coverage was restricted to 8--13\um\ by severe atmospheric absorption. \textit{JWST} spectroscopy of SN\,2024pxl and SN\,2024vjm, two peculiar Type Iax supernovae (SN\,Iax; \citealt{Li2003, Foley2013}), provided the first complete MIR view (2.5--28\um) of a thermonuclear SN during the early transitional phase ($\sim$+10--40\,days post-maximum; \citealt{Kwok2025a}). These observations showed that forbidden emission can emerge early in the MIR, although SN\,Iax differ from normal \snia\ and therefore cannot establish when MIR forbidden emission should appear in normal \snia. Before this work, no MIR spectrum of a normal \snia\ had been obtained at or near maximum light, leaving its MIR photospheric and early transitional phases unexplored.

Here, we present \textit{JWST} spectroscopy of the nearby normal \snia\,2025rbs in NGC\,7331 from 2.9--14\um\ at $+1$, $+23$, and $+$84~days post \textit{B}-band maximum through program \textit{JWST}-GO-9255 \citep{Kwok2025_9255}. This constitutes the earliest MIR spectroscopic sequence of an \snia\ to date, including the first maximum-light spectrum. Combined with contemporaneous ground-based optical and NIR spectroscopy, our panchromatic sequence provides nearly continuous coverage from 0.4--14\um\ across the photospheric, transitional, and early nebular phases. We use these observations to characterize the MIR spectral evolution of a normal \snia\ for the first time, identify the ions contributing to the complex transitional-phase spectra, and place new constraints on the ejecta structure and explosion physics of \rbs.

\begin{figure*}
    \centering
    \includegraphics[width=\linewidth]{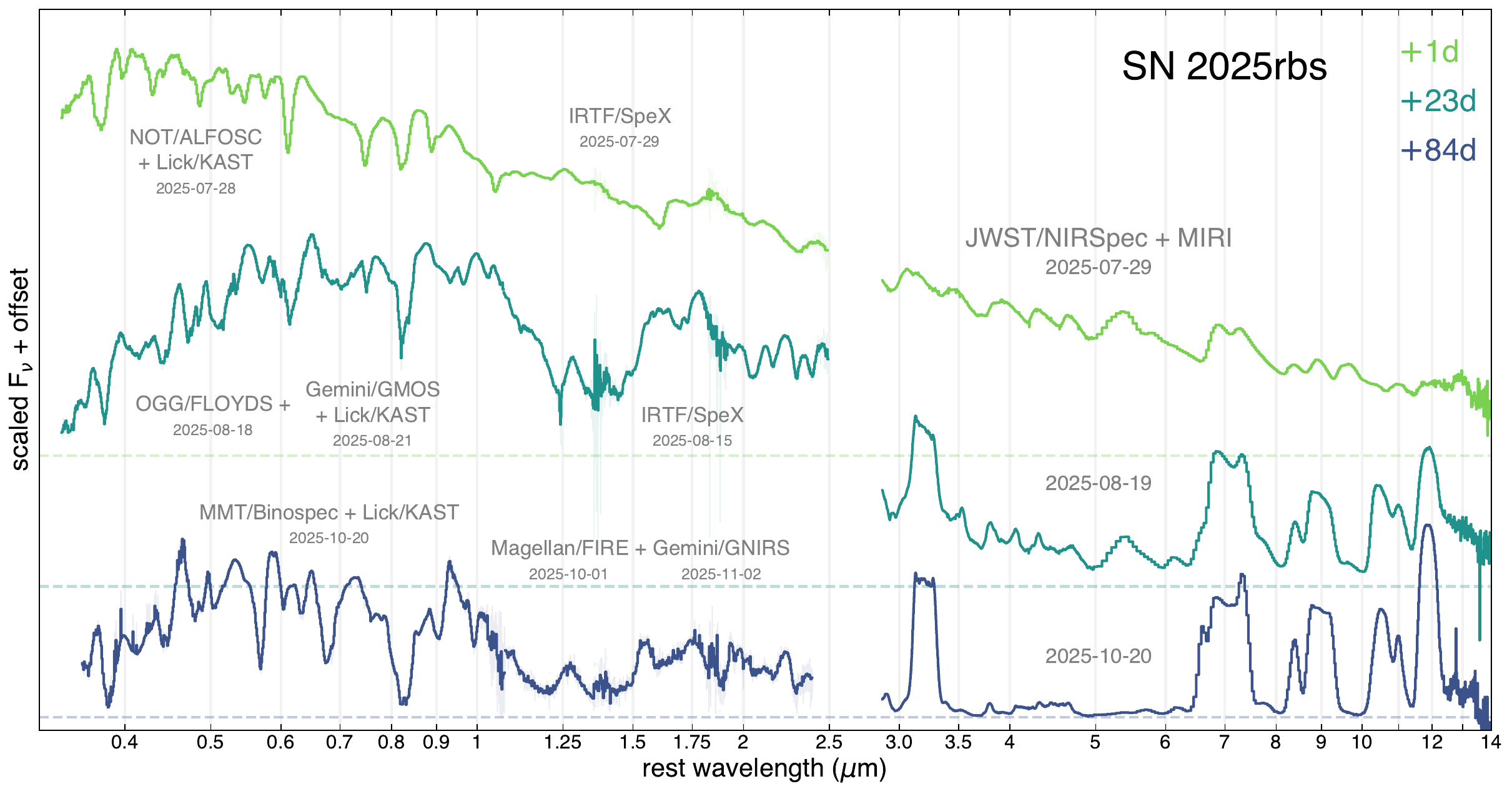}
    \caption{Panchromatic spectra of \rbs\ at $+$1, $+$23, and $+$84\,days post-\textit{B}$_{\rm max}$. Flux density is shown in an arcsinh scaling for display purposes, and the epochs are offset for clarity, with the zero-point marked by the dashed lines. Each spectrum is a combination of {\it JWST} NIR and MIR spectra with ground-based optical and NIR spectra at similar phases.}
    \label{fig:panchromatic}
\end{figure*}

\section{Observations \label{sec:obs}}

\rbs\ ($\alpha = 22^{\mathrm{hr}} 37^{\mathrm{m}} 03^{\mathrm{s}}_{^{\centerdot}}658$, $\delta = +34^\circ25' 07\farcs98$, J2000) in NGC\,7331 was discovered and reported to the Transient Name Server\footnote{\url{https://www.wis-tns.org/object/2025rbs}} (TNS) by the GOTO Team on 14 July 2025 at 10:23:53 \citep[UTC dates are used throughout;][]{ONeill2025} and was classified by the Global Supernova Project (GSP) as an SN\,Ia on 14~July~2025 at 14:33:55 \citep{Andrews2025}.

\subsection{JWST Data}

\rbs\ was observed with the \textit{JWST} Near-Infrared Spectrograph (NIRSpec) Fixed Slit (FS) G395M grating \citep{Jakobsen2022,Birkmann2022,Rigby2022} and the Mid-Infrared Instrument (MIRI) Low Resolution Spectrograph \citep[LRS;][]{Kendrew2015,Kendrew2016,Rigby2022} on 29~July~2025, 19~August~2025, and 20~October~2025. With \textit{B}-band maximum on 28 July 2025 (see \autoref{sec:light_curve}), these correspond to phases of $+$1, $+$23, and $+$84\,days.

Our observations used the NIRSpec/FS with the S200A1 slit and the G395M grating (resolving power $R\approx1000$), and the MIRI/LRS slit with the P750L disperser ($R\approx100$), spanning 2.9--14\um. We choose to transition from the NIRSpec spectrum to the MIRI spectrum at 5\um, as NIRSpec coverage ends and MIRI coverage begins near this wavelength.
The NIRSpec and MIRI data were reduced using the automatic \textit{JWST} pipeline, available on the Mikulski Archive for Space Telescopes (MAST)\footnote{All \textit{JWST} data are publicly available on MAST at \dataset[DOI: 10.17909/e7tz-hd05]{https://doi.org/10.17909/e7tz-hd05}}, and are presented in \autoref{fig:panchromatic}. The MIRI/LRS slit-mode wavelength calibration is known to be uncertain at the shortest wavelengths, so we apply the additional LRS wavelength-calibration correction from \citet[][see \href{https://iopscience.iop.org/article/10.3847/2041-8213/adf062\#apjladf062app1}{Appendix A}]{Kwok2025a} to all epochs of LRS data of \rbs. 


\subsection{Ground-based Optical and NIR Spectroscopy}

Optical spectra were obtained at nearly contemporaneous phases as the \textit{JWST} observations with the Nordic Optical Telescope's (NOT) Alhambra Faint Object Spectrograph and Camera (ALFOSC), the Lick Observatory's Kast double spectrograph (Lick/Kast), the MMT Observatory's Binospec instrument (MMT/Binospec), the Las Cumbres Observatory's FLOYDS spectrograph, and the Gemini North telescope's Multi-Object Spectrograph (Gemini/GMOS).

The NOT spectra were reduced using a custom data-reduction pipeline based on \texttt{pypeit} \citep{pypeit:joss_arXiv, pypeit:zenodo, pypeit:joss_pub}\footnote{\url{https://pypeit.readthedocs.io/en/latest/}}, which is available on GitHub\footnote{\url{ https://github.com/steveschulze/NOT_DRP}}.

We observed \rbs\ with the Kast dual-beam spectrograph \citep{KAST} on the 3\,m Shane telescope at Lick Observatory. Spectra were obtained on 2025 July 28.397 and 28.405 ($+$0.37 and $+$0.38\,days), which we combine via inverse-variance weighting, along with spectra on each of 2025 Aug 21.4 ($+$24.4\,days) and 2025 Oct 21.2 ($+$85.2\,days), contemporaneous with all three \textit{JWST} epochs. The spectra were reduced following standard procedures\footnote{\url{https://github.com/msiebert1/UCSC_spectral_pipeline}}\textsuperscript{,}\footnote{\url{https://github.com/ishivvers/TheKastShiv}} as outlined by \citet{Silverman2012} and \citet{Siebert2019}.

We observed SN\,2025rbs with the FLOYDS spectrograph \citep{Brown2013} on the 2\,m Faulkes Telescope North (FTN) at Haleakala Observatory, part of the Las Cumbres Observatory Global Telescope Network \citep{Brown2013}, through the Global Supernova Project (GSP) collaboration, on 2025 Aug 18.3 ($+$21.3\,days). The spectrum was reduced with the \texttt{floydsspec}\footnote{\url{https://www.authorea.com/users/598/articles/6566}} pipeline using standard reduction techniques.

We observed SN\,2025rbs with the Gemini Multi-Object Spectrograph at Gemini North (GMOS-N; \citealt{2004PASP..116..425H}) on 2025 Aug 21.5 ($+$24.5\,days). The data were reduced using Gemini IRAF\footnote{IRAF is distributed by the National Optical Astronomy Observatories, operated by the Association of Universities for Research in Astronomy, Inc., under a cooperative agreement with the U.S. National Science Foundation (NSF).} packages.

We observed SN\,2025rbs with the Binospec instrument \citep{Fabricant+2019} on the 6.5\,m MMT telescope at the Fred Lawrence Whipple Observatory on 2025 Oct 20.3 ($+$84.3\,days). The 270 lines mm$^{-1}$ grating centered at 6500 \AA~yielded broad optical coverage with $R\approx1340$. The reduction was completed with standard procedures implemented in \texttt{pypeit} \citep{pypeit:joss_arXiv, pypeit:zenodo, pypeit:joss_pub}.

NIR spectra were obtained with the NASA Infrared Telescope Facility (IRTF) SpeX spectrograph \citep{Rayner03} on 2025 July 29.5 ($+$1\,day), contemporaneous with the first \textit{JWST} spectrum, and on 2025 Aug 15.5 ($+$18\,days), similar in phase to the second \textit{JWST} spectrum, by program 2025B079 (PI W. B. Hoogendam) as part of the Hawaii Infrared Supernova Study (HISS; \citealp{Medler25_HISS, Hoogendam25_epr, Hoogendam25_pxl, Medler2025}). The spectra were reduced using standard IRTF \texttt{Spextool} procedures with an A0\,V standard star to correct for telluric features and flux calibrate the data \citep{Vacca_2003, Cushing04}.

For the third epoch of \textit{JWST} observations, we combine a NIR spectrum from the Magellan/Baade Folded-port InfraRed Echellette (FIRE) spectrograph on 2025 Oct 2.1 ($+$66.1\,days; PI: C.~Liu) with a NIR spectrum from the Gemini North GNIRS spectrograph on 2025 Nov 2.2 ($+$97.2\,days; PI L.~A.~Kwok). 
The FIRE spectrum was reduced with \texttt{pypeit} \citep{pypeit:joss_arXiv, pypeit:zenodo, pypeit:joss_pub}. The GNIRS XD spectrum was reduced using the GNIRS module in the Gemini DRAGONS reduction pipeline \citep{Labrie2023, Simpson2026}. As these spectra were taken significantly before and after the {\it JWST} observation date, we account for the evolution of the SN between epochs by interpolating the flux to the epoch of the \textit{JWST}/NIRSpec observation ($+84.6$\,days). The interpolation is performed linearly in magnitude ($\log F_\lambda$) as a function of time, with each spectrum weighted according to its fractional phase separation from the target epoch (giving the FIRE and GNIRS spectra weights of $~$0.4 and $~$0.6, respectively).

\subsection{Optical Photometry \label{sec:light_curve}}

Optical photometry of \rbs\ was obtained through the GSP collaboration using the Las Cumbres Observatory (\citealt{Brown2013}) 0.4~m and 1~m telescopes in the $BVgri$ bands. Preprocessing, including bias correction and flat fielding, was handled by the \texttt{BANZAI} pipeline \citep{McCully2018}. Additional photometry was obtained with the Katzman Automatic Imaging Telescope (KAIT) and the Nickel 1\,m telescope at the Lick Observatory in the \textit{BVRI} and {\it Clear} bands. The images were calibrated using bias and sky flat-field frames following standard procedures. Point-spread-function (PSF) photometry was performed and several nearby stars were chosen from the Pan-STARRS1 \citep{Flewelling16} catalog for calibration; their magnitudes were transformed into Landolt \citep{landolt92} magnitudes using the empirical prescription presented by Eq.~6 of \citet{Tonry2012}.

The early discovery and rapid classification of \rbs\ enabled excellent early photometric coverage, and a detailed analysis of the light curve will be presented in future work. Here, we use the optical light curves only to establish the distance, extinction, and epoch of maximum light for the spectral analysis presented in this work.

We fit the light curves using \texttt{BayeSN}, a hierarchical Bayesian SN\,Ia light-curve model \citep{mandel_hierarchical_2022, Grayling2024}, which models host-galaxy dust extinction separately from intrinsic spectral energy distribution (SED) variations. This is particularly important given the location of \rbs\ near the nucleus of NGC~7331. Using the model trained by \citet{ward_bayesn_2023}, we obtain a distance modulus of $\mu = 30.73 \pm 0.12$\,mag ($14.0 \pm 0.8$\,Mpc), in good agreement with the Cepheid distance of \citet[][see their Table 11]{Freedman2001} of $\mu = 30.84 \pm 0.09$\,mag ($14.5 \pm 0.6$\,Mpc). We adopt the more precise Cepheid distance of $14.5$\,Mpc throughout. The fit also yields a host-galaxy extinction of $A_V = 0.35 \pm 0.07$\,mag (with $R_V = 2.39 \pm 0.45$), a light-curve decline rate of $\Delta m_{15}(B) = 1.47 \pm 0.01$\,mag, and a $B$-band maximum of MJD~$= 60884.03 \pm 0.05$, corresponding to 2025 July 28. The resulting interstellar extinction of $E(B-V) = 0.15 \pm 0.04$\,mag from our light-curve fitting is consistent with the total line-of-sight reddening of $E (B-V) = 0.21 \pm 0.10$\,mag independently estimated from the observed diffuse interstellar band (DIB) at 5780 \AA\ in a high-resolution spectrum of \rbs\ \citep{Ravi2026}. Adopting the Cepheid distance, this corresponds to a host-extinction-corrected peak absolute magnitude of $M_B = -19.13 \pm 0.11$\,mag. With $\Delta m_{15}(B) = 1.47$\,mag and $M_B = -19.13$\,mag, \rbs\ falls on the normal-but-fast-declining end of the Phillips relation \citep{Phillips1993}, noticeably faster declining and modestly fainter than the canonical SN\,2011fe.

As an independent check, we separately fit ground-based $BVRI$ photometry from the KAIT/Nickel telescopes. The two fits agree well on all extinction- and distance-sensitive parameters ($A_V$, $R_V$, $\mu$, $\Delta m_{15}(B)$, $M_B$; all consistent to $<1\sigma$). We fit each set of data independently, as combining them results in a poor fit, likely owing to cross-calibration uncertainty. We prefer the fit using Las Cumbres Observatory photometry because the rise to peak is better sampled. The two fits nonetheless yield distance moduli in good agreement ($\mu = 30.73$ vs. $30.82$\,mag, $\Delta\mu = 0.09$\,mag, within $0.5\sigma$), supporting the robustness of the derived distance independent of which photometric dataset is adopted.

We also estimate the rise time of \rbs\ by fitting its early-time light curves with power-law rise models, following the methodology of \citet{Liu2026a}. We restrict the fit to the Las Cumbres Observatory $g$- and $r$-band observations, allowing us to use population-level information on the power-law parameters \citep{Liu2026b} inferred from the ZTF SN\,Ia Data Release 2 (DR2) sample \citep{Rigault2025}. Specifically, as suggested by \citet{Liu2026a}, we retain the population correlation structure through a Gaussian copula, while adopting broad uniform marginal priors for the rise time, $t_{\mathrm{rise}}$, and the power-law indices, $\alpha_g$ and $\alpha_r$, to limit shrinkage toward the population means. 
Because all Las Cumbres Observatory observations were obtained after the onset of \rbs, we supplement them with public ZTF $g$- and $r$-band forced-photometry nondetections extending back to 100\,days before $B$-band maximum. The light curves were fit only up to the epoch at which they reach 30\% of the estimated maximum flux in each filter. We infer a rise time of $16.6\pm0.3$\,days from the estimated first-light epoch\footnote{This empirical estimate of first optical brightening epoch is not necessarily a measurement of the explosion epoch.} to $B$-band maximum, making \rbs\ a relatively rapid riser compared with the ZTF DR2 sample, which has a mean rise time of 18.55\,days \citep{Liu2026b}, and with several earlier SN\,Ia samples \citep[e.g.,][]{Hayden2010,Firth2015,Miller2020}. We also infer steeper-than-average rise indices in both $g$ ($\alpha_g=3.08\pm0.14$) and $r$ ($\alpha_r=2.41\pm0.15$) relative to the ZTF DR2 SN\,Ia sample, for which the corresponding mean values are $\alpha_g=2.29$ and $\alpha_r=2.10$. The substantial difference, $\alpha_g-\alpha_r=0.67$, further indicates significant red-to-blue color evolution during the early rise.

\begin{figure}
    \centering
    \includegraphics[width=\linewidth]{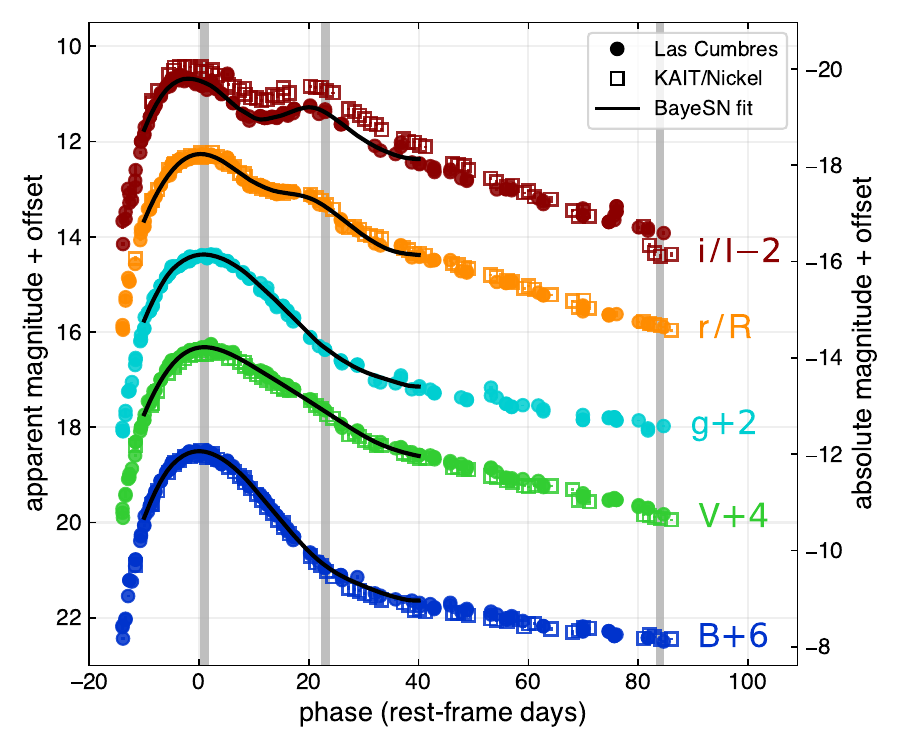}
    \caption{Light curve of \rbs\ from Las Cumbres Observatory (circles; $BVgri$) and KAIT/Nickel (squares; $BVRI$), offset for clarity. Rest-frame phase is with respect to \textit{B}-band maximum (MJD 60884.0). The BayeSN fit to the Las Cumbres Observatory data between $-$10 and $+$40\,days is shown in black. The \textit{JWST} observation epochs are marked by vertical gray lines.}
    \label{fig:light_curve}
\end{figure}

\section{Spectral Analysis \label{subsec:tables}}

\begin{figure*}
    \centering
    \includegraphics[width=\linewidth]{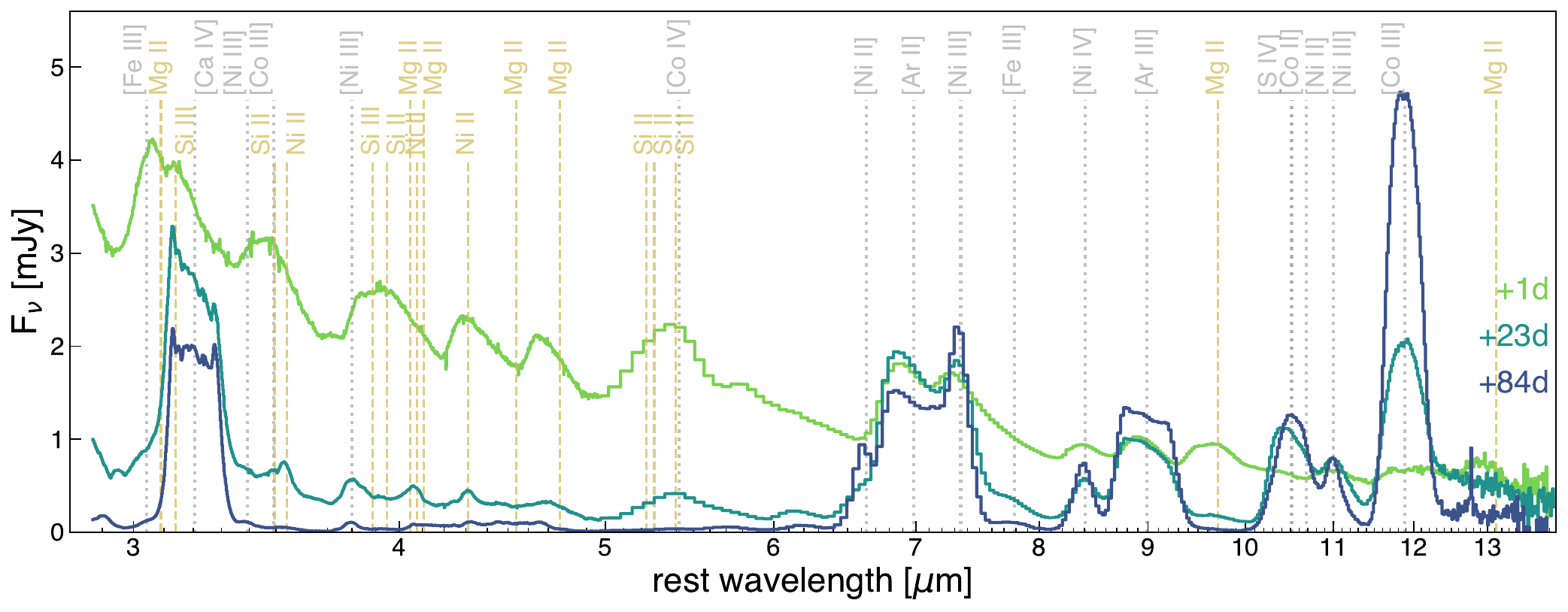}
    \caption{Line identifications from 2.9--14\um\ for \rbs\ at $+$1 (green), $+$23 (teal), and $+$84\,days (indigo). Permitted transitions are marked by yellow dashed lines, and forbidden transitions are marked by gray dotted lines. Only the most dominant contributors to each feature are labeled (see \autoref{sec:appendix}).}
    \label{fig:jwst_spec}
\end{figure*}

We directly integrate the pseudobolometric luminosity of \rbs\ in the range 0.4--14\um\ at each \textit{JWST} epoch and summarize the fractional contribution of optical (0.4--1.0\um), NIR (1.0--2.5\um), and MIR (2.8--14\um) wavelengths in \autoref{tab:mir_fraction}. The optical dominates the bolometric budget throughout, contributing $\sim$90\% of the total at maximum light and $\sim$75--80\% at the later epochs, while the NIR contributes $\sim$10--20\%, with the largest contribution at $+23$\,days, consistent with the timing of the second NIR light-curve peak. The MIR contribution grows substantially with phase, from $<$1\% at $+$1\,day to $\sim$7\% by $+$84\,days. While the MIR flux is negligible to the bolometric budget at maximum light, the spectrum already reveals features probing physical conditions in the ejecta inaccessible at optical and NIR wavelengths.

We note that the remaining $\lesssim$1\% at each epoch falls in the 2.5--2.8\um\ gap between the ground-based NIR and \textit{JWST} coverage and is not attributed to any of the three bands above. Additionally, the $+$23 and $+$84\,day NIR contributions are derived from data not strictly contemporaneous with the \textit{JWST} epoch. The NIR contribution is likely overestimated at $+$23\,d, since the IRTF spectrum was obtained 4\,days prior to the \textit{JWST} observation, and is likewise uncertain at $+$84\,d, where it relies on phase interpolation between the FIRE and GNIRS spectra obtained $-18.5$ and $+12.6$\,d relative to the \textit{JWST} observation, respectively. The $+$1\,day values, by contrast, are robust, as the optical and NIR spectra were obtained within a day of the \textit{JWST} observation.

\begin{table}
\centering
\caption{Fractional contribution of optical, NIR, and MIR wavelengths to the pseudobolometric flux, integrated between 0.4 to 14\um, of \rbs\ at each \textit{JWST} epoch. \label{tab:mir_fraction}}
\label{tab:mir_fraction}
\begin{tabular}{lcccc}
\toprule
Phase & Optical & NIR & MIR & \\
(days) & (0.4--1.0\,\micron) & (1.0--2.5\,\micron) & (2.8--14\,\micron) & \\
\midrule
$+$1   & 87.5\% & 11.3\% & 0.8\% & \\
$+$23  & $\sim$76\% & $\sim$22\%\tablenotemark{a} & $\sim$1\% & \\
$+$84  & $\sim$78\% & $\sim$14\%\tablenotemark{b} & $\sim$7\% & \\
\bottomrule
\end{tabular}
{\raggedright
\tablenotetext{a}{Likely overestimated; NIR spectrum obtained 4\,d before the \textit{JWST} epoch.}
\tablenotetext{b}{Uncertain due to phase interpolation between NIR spectra obtained $19$\,d before and $13$\,d after the \textit{JWST} epoch.}
\par}
\end{table}

\subsection{Line Identification \label{sec:lineIDs}}

At $+$84~days, \rbs\ displays the same major forbidden-line emission in the MIR as seen in previous studies of normal SN\,Ia \citep{Gerardy2007, Kwok2023, DerKacy2023, Blondin2023, Ashall2024, Kwok2026}, with identified lines marked in \autoref{fig:jwst_spec}. The earlier epochs, particularly at $+$1\,day, host a mixture of permitted and forbidden lines, many of which have not been previously accessible owing to the lack of MIR observations at these early phases. To aid in line identification at these transitional epochs, when the ejecta are still partially optically thick, we use model line lists from one-dimensional (1D) CMFGEN \citep{Hillier2012} radiative-transfer models based on the angle-averaged outputs of the N100 delayed-detonation model of \cite{Seitenzahl2013}, computed at $\sim$20, 40, and 100\,days post-explosion. These models are described further in \autoref{sec:models}.

At maximum light ($+1$~day), the \textit{JWST} spectrum exhibits a prominent continuum superposed with broad features. Between 3\um\ and 6\um, we identify prominent permitted lines from \ion{Si}{2}, \ion{Si}{3}, and \ion{Ni}{2}, with potential contributions from \ion{Mg}{2} and various IGE lines (see \autoref{sec:models}). These features lack pronounced absorption components, though blending may obscure such structure. Redward of 6.5\um, forbidden lines are already emerging: notably, [\ion{Ar}{2}]\,6.98\um, [\ion{Ni}{3}]\,7.35\um, [\ion{Ni}{4}]\,8.41\um, and [\ion{Ar}{3}]\,8.99\um. We also detect features near 4.7, 9.6, and 13.0\um\ that disappear in subsequent epochs (\autoref{fig:neb_evolve}); we identify these as \ion{Mg}{2}\,4.76, 9.71, and 13.12\um\ (see \autoref{sec:models}). Several forbidden lines may also contribute to the features in the range 3--6\um, including [\ion{Ca}{4}]\,3.21\um\ and [\ion{Ni}{3}]\,3.39, 3.80\um. These forbidden lines become more isolated at later epochs as the surrounding permitted lines fade. The permitted and forbidden transitions marked in \autoref{fig:jwst_spec} are given in \autoref{sec:appendix}.

We note that the $+1$~day spectrum shows little [\ion{Co}{2}] and only a weak potential [\ion{Co}{3}] feature (see also \autoref{fig:neb_evolve}). There may be weak permitted lines, continuum, or pseudocontinuum contaminating these regions at this phase, so it is unclear whether the Co emission is genuinely weak or partially masked; indeed, our radiative-transfer model (\autoref{sec:models}) predicts a [\ion{Co}{3}]\,11.89\um\ line at this epoch, though substantially weaker than the [\ion{Ni}{3}]\,7.35\um\ feature. In the observations, it appears that [\ion{Co}{3}] is much less prominent than [\ion{Ni}{3}]. By maximum light, $\gtrsim$85\% of the $^{56}$Ni has decayed to $^{56}$Co, so the weak Co emission is not an abundance issue. Furthermore, while the weakness of [\ion{Co}{2}] is consistent with the lack of [\ion{Ni}{2}] and can be explained by high ionization and high temperature at this phase, we suggest the disparity between [\ion{Ni}{3}] and [\ion{Co}{3}] may arise from a difference in critical density. 

Assuming they form in similar layers (i.e., the same temperature), the critical density for a level-two transition in [\ion{Co}{3}] is $\sim$5--6 times lower than the same level for [\ion{Ni}{3}], meaning [\ion{Co}{3}] emission could have a delayed emergence as the densities continue to drop below its critical density. In addition, or alternatively, since Ni and Co occupy somewhat separate regions of the ejecta (see \autoref{sec:neb_lines}), with Ni concentrated in denser inner regions, recombination can be enhanced there early relative to Co, which might remain at a higher ionization state early (e.g., [\ion{Co}{4}]) before recombining to [\ion{Co}{3}] later.

\subsection{MIR Spectral Evolution \label{sec:mir_evolution}}

\begin{figure*}
    \centering
    \includegraphics[width=\linewidth]{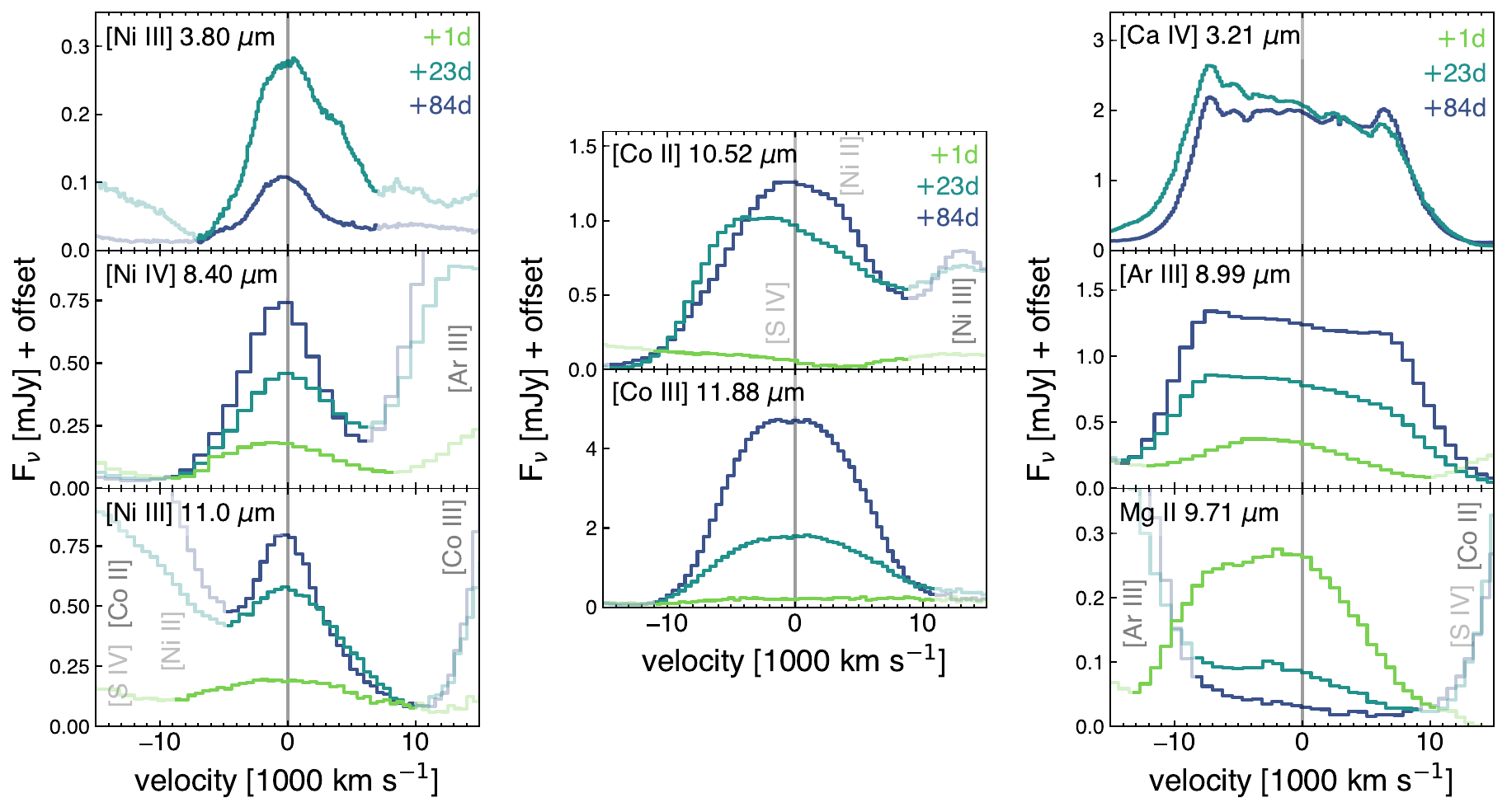}
    \caption{Evolution of relatively isolated MIR forbidden lines in \rbs, together with the candidate permitted-like \ion{Mg}{2}\,9.71\um\ feature, at $+$1 (green), $+$23 (teal), and $+$84\,days (indigo). Lower-opacity portions of each profile indicate wavelength regions contaminated by nearby transitions, which are labeled; fainter gray labels denote weaker contributors. To account for the changing continuum level, the $+$1 and $+$23\,day spectra are shifted vertically so that the local continuum around each transition matches that of the $+$84\,day spectrum. The $+$1\,day profiles of [\ion{Ni}{3}]\,3.80\um\ and [\ion{Ca}{4}]\,3.21\um\ are omitted because these lines cannot be isolated from strong surrounding permitted emission at this epoch.}
    \label{fig:neb_evolve}
\end{figure*}

The \textit{JWST} spectra of \rbs\ show rapid evolution (\autoref{fig:jwst_spec}). The continuum fades significantly between $+$1 and $+$23\,days, and is absent by $+$84\,days, indicating that the ejecta are fully nebular in the MIR by this phase. As the temperature and density drop, the permitted lines (e.g., \ion{Si}{2}, \ion{Si}{3}, and \ion{Ni}{2}) likewise fade from $+1$ to $+23$\,days and are not detected at $+84$\,days, while forbidden emission lines strengthen. The candidate \ion{Mg}{2}\,4.76, 9.71, and 13.12\um\ lines, however, follow the evolution of the permitted lines rather than the forbidden ones, disappearing by $+$23\,days (\autoref{fig:neb_evolve}). 

In the CMFGEN atomic data used to generate our model line lists \citep[see][]{Blondin2023}, the upper levels of these transitions combine multiple high-$\ell$ (orbital angular momentum) states, a simplification adopted for computational tractability. These combined states do not have a uniquely defined parity, so the transitions cannot be classified straightforwardly as forbidden or permitted. Their radiative rates are nevertheless comparable to those of permitted transitions (e.g., $A\approx 4\times10^6$\,s$^{-1}$ for the 4.76\um\ transition), consistent with their rapid fading alongside the permitted lines and continuum. Given these properties, we refer to these combined-state \ion{Mg}{2} transitions as ``permitted-like'' throughout this work. The large statistical weights (i.e., from multiplicity) and high excitation energies of the combined states suggest that they are populated primarily through recombination cascades. The use of combined states is therefore not expected to artificially enhance the integrated line fluxes, although it may produce profiles that are somewhat narrower than if the individual states were treated separately, since the emission would then be distributed over a slightly broader wavelength range.

\autoref{fig:neb_evolve} shows the evolution of several key MIR lines through the nebular transition. At $+$1\,day, the [\ion{Ar}{3}]\,8.99\um\ line is largely optically thick. By $+$23\,days, the blue side of the profile has developed a flat top while the red side remains partially obscured; by $+$84\,days, the red side has likewise sharpened into a clean edge, indicating that the ejecta are fully optically thin in the MIR by this epoch. The tilt of the flat-topped [\ion{Ar}{3}] profile---and of [\ion{Ar}{2}]\,6.98\um, though blending with nearby Ni lines makes this less clear---is consistent with MIR observations of other normal SN\,Ia \citep{Kwok2023, DerKacy2023, Ashall2024, Kwok2026}, where variations in the tilt may reflect differences in viewing angle for mildly off-center explosions.

The [\ion{Ca}{4}]\,3.21\um\ line, covered by the medium-resolution NIRSpec G395M grating, is broadly consistent with [\ion{Ar}{3}] in width and profile shape, but additionally resolves substructure within the ejecta. While some excess blue flux from fading continuum and permitted lines persists in the $+23$\,day spectrum, the same fluctuations in the flat-topped profile appear at both $+$23 and $+$84\,days, confirming that this substructure is real. Notably, different substructure signatures are seen in the [\ion{Ca}{4}] profiles of the normal SN\,Ia 2022aaiq and 2024gy \citep{Kwok2026}; were the fluctuations caused by blending with weaker lines, their positions and relative strengths would be expected to be consistent across objects (see \autoref{sec:ca_substructure}).

The isolated, permitted-like \ion{Mg}{2}\,9.71\um\ line evolves quickly and exhibits a markedly blueshifted, boxy shape at $+$1\,day. At this phase, the ejecta are still largely optically thick, and obscuration of the receding ejecta may suppress the red side of the profile and produce the apparent blueshift. The blue side of the \ion{Mg}{2} line extends to substantially higher velocity than [\ion{Ar}{3}] at $+$1\,day, reaching an extent comparable to that of [\ion{Ar}{3}] at $+$84\,days. As discussed above, the high radiative rate of the \ion{Mg}{2}\,9.71\um\ transition allows the excited state to decay efficiently once populated. Its highly excited upper states are likely populated through recombination cascades. The rapid disappearance of the line may therefore reflect declining recombination flow through these states, evolution of the overall \ion{Mg}{2}/\ion{Mg}{3} ionization balance, and the weakening radiation field as the ejecta expand and cool.

Alternatively, or in addition, the extended \ion{Mg}{2} profile may indicate that Mg resides farther out in the ejecta than Ar and the \ion{Mg}{2} transitions become optically thin earlier owing to the lower density and diluted radiation field in those layers. Such an ejecta structure would be qualitatively consistent with a stratified explosion, including delayed- or double-detonation scenarios in which carbon-burning products such as Mg occupy the outermost layers. The velocity extent and separation of the Mg- and Ar-emitting regions could help differentiate between explosion models, although the observed profiles also depend on ionization, excitation, and optical depth, requiring radiative-transfer calculations to distinguish between possibilities. Because the candidate MIR \ion{Mg}{2} lines fade by $+$23\,days, observations within the first few weeks post-maximum provide the only opportunity to study them. We cannot rule out a misidentification, but we find no compelling alternative line candidates near 9.7\um, and the correspondence of multiple predicted \ion{Mg}{2} transitions with otherwise unexplained features strengthens this identification (see \autoref{sec:models}).

\begin{figure*}
    \centering    
    \includegraphics[width=\linewidth]{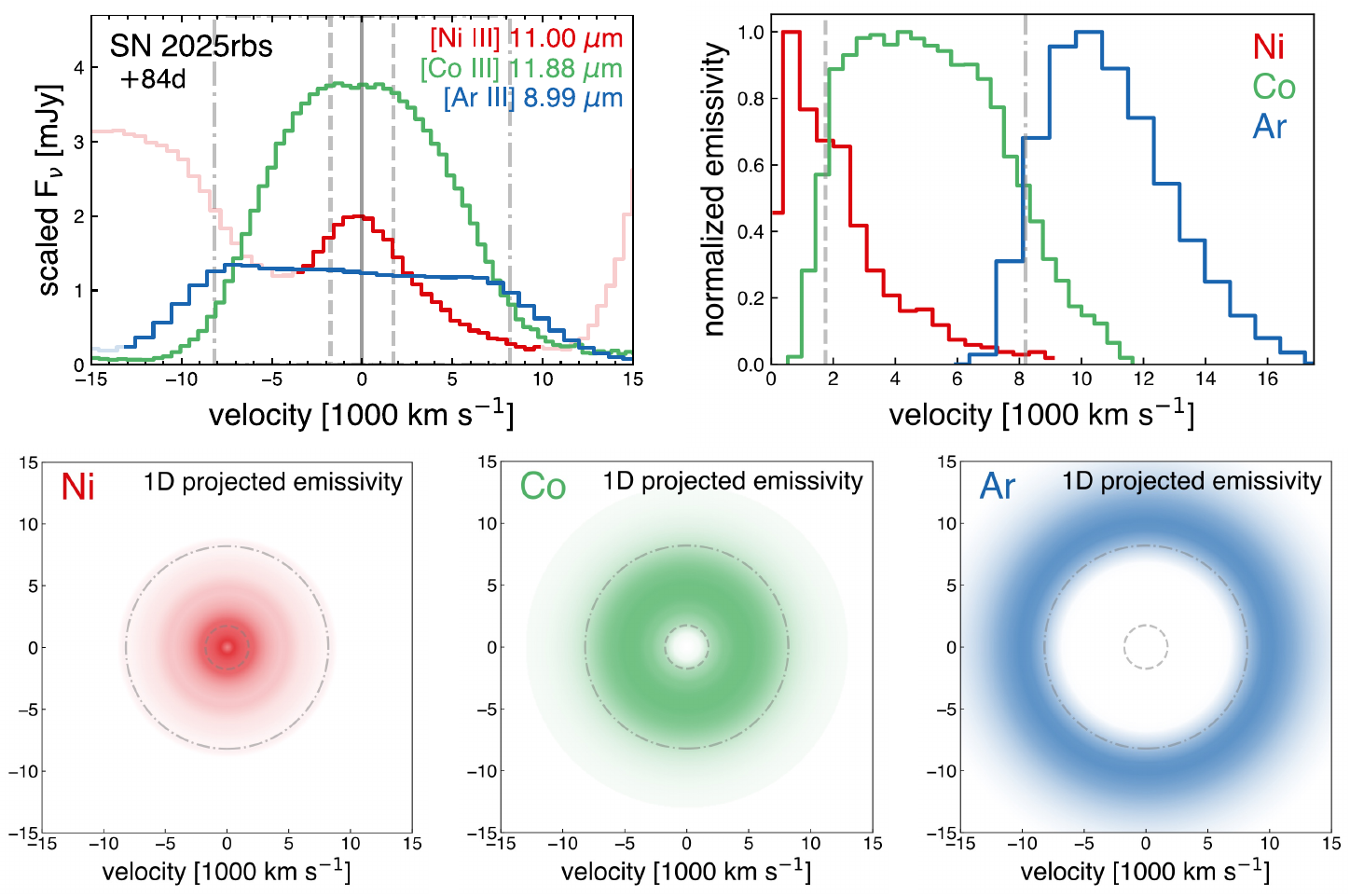}
    \caption{\textit{Top left}: [\ion{Ni}{3}]\,11.00\um\ (red), [\ion{Co}{3}]\,11.88\um\ (green), and [\ion{Ar}{3}]\,8.99\um\ (blue) line profiles of \rbs\ at $+$84\,days. The dashed gray lines at $\pm$1750\kms\ represent the inner edge of the Co flat top, and the dash-dotted gray lines at $\pm$8200\kms\ represent the inner Ar edge; these lines are consistent across all panels. \textit{Top right}: 1D normalized emissivity profiles generated from the red side of the [\ion{Ni}{3}]\,11.00\um, [\ion{Co}{3}]\,11.88\um, and [\ion{Ar}{3}]\,8.99\um\ line profiles. \textit{Bottom}: 1D projected emissivity for Ni (\textit{left}) tracing the stable IGEs, Co (\textit{center}) tracing the radioactive IGEs, and Ar (\textit{right}) tracing the IMEs. These plots are illustrative and generated from the 1D emissivities in the top-right panel, assuming radial symmetry. Minor radial fluctuations in the emissivity structure (e.g., the bump at $\sim4700$\kms\ in Ni) should not be overinterpreted, as they may be affected by weak nearby lines.} 
    \label{fig:emissivity}
\end{figure*}

At maximum light, we do not clearly detect [\ion{Ni}{2}]\,6.636\um, [\ion{Co}{2}]\,10.52\um, or [\ion{Co}{3}]\,11.88\um. The subsequent emergence and strengthening of these lines over time suggests a decrease in the mean ionization state of the IGE species as the ejecta expand and cool. In contrast, the clearly detected [\ion{Ni}{3}]\,7.35\um\ line shows an extended red wing at $+$1\,day that diminishes substantially by $+$23\,days, which might reflect a contribution from radioactive $^{56}$Ni located in the outer ejecta at this phase.

By $+$23\,days, [\ion{Co}{2}]\,10.52\um\ and [\ion{Co}{3}]\,11.88\um\ have emerged and display broader, less peaked profiles than the Ni lines. The origin of this difference is key to interpreting the explosion. A flattened line core could in principle reflect that the innermost layers have not yet become fully optically thin; however, the absence of a photospheric continuum, the fully nebular [\ion{Ar}{3}]\,8.99\um\ profile, and the peaked Ni lines collectively demonstrate that the ejecta are optically thin by $+$84\,days, ruling out an opacity origin. We therefore attribute the flattened top of the [\ion{Co}{3}]\,11.88\um\ profile to indicate the absence of emission at velocities $\lesssim$2000\kms, which we attribute to the underlying spatial distribution of radioactive $^{56}$Co relative to stable Ni. This indicates that the central region is dominated by stable IGEs, with radioactive material concentrated at higher velocities.

\section{MIR Nebular-Line Analysis \label{sec:neb_lines}}

In the following analysis, we assume that the MIR lines at $+$84\,days are optically thin. This assumption is supported by the emergence of the sharp transition between the plateau and red wing of the [\ion{Ar}{3}]\,8.99\um\ profile, the disappearance of the underlying MIR continuum at this phase, and the dominance of forbidden-line emission. We analyze the nebular $+$84\,day spectrum following the fitting procedures outlined in more detail in previous work \citep{Kwok2023, Kwok2024, Kwok2025a, Kwok2026} and the line-inversion procedure from \cite{Kwok2026}. The projected emissivities shown in \autoref{fig:emissivity} are generated directly from the line profiles using
\begin{equation}\label{eq:1}
    j(v) \propto \frac{1}{v}\,\frac{{\rm d}F_\nu(v)}{{\rm d}v}
\end{equation}
\citep{Fransson1989, Jerkstrand2017, Kwok2026}. Equation~\ref{eq:1} depends on the 3D velocity; however, lacking this, we use the line-of-sight velocity as a proxy, which introduces the assumption of symmetry about the line of sight. The red side of each 1D emissivity profile is shown in \autoref{fig:emissivity} (\textit{top-right}) and are visualized in the bottom panels, assuming radial symmetry. These visualizations are meant to be illustrative, and not a true reconstruction of the projected 3D structure.

\begin{table}
\footnotesize
\setlength{\tabcolsep}{5pt}
\caption{Flux-space stratification metrics for \rbs, derived from the
volume-weighted, normalized emissivity from \autoref{fig:emissivity}. Values
are bootstrap median with 16th--84th percentile uncertainty intervals,
including the instrumental-resolution systematic added in quadrature (see
text).}
\label{tab:overlap_metrics}
\begin{tabular}{l c c c}
\toprule
\multicolumn{4}{l}{\textit{Flux-weighted velocity quantiles [$10^3$ km s$^{-1}$]}} \\
Line & $v_{10}$ & $v_{50}$ & $v_{90}$ \\
\cmidrule(lr){1-4}
{[}\ion{Ni}{3}{]} 11.00\,$\mu$m & $2.2^{+0.2}_{-0.2}$ & $5.0^{+0.4}_{-0.4}$ & $8.3^{+0.6}_{-0.8}$ \\
{[}\ion{Co}{3}{]} 11.89\,$\mu$m & $4.1^{+0.1}_{-0.1}$ & $7.0^{+0.1}_{-0.1}$ & $9.7^{+0.1}_{-0.1}$ \\
{[}\ion{Ar}{3}{]} 8.99\,$\mu$m & $8.9^{+0.3}_{-0.3}$ & $11.6^{+0.2}_{-0.1}$ & $14.9^{+0.8}_{-0.5}$ \\
\specialrule{\heavyrulewidth}{3pt}{3pt}
\multicolumn{4}{l}{\textit{Pairwise flux overlap, OVL$(A,B)$}} \\
\multicolumn{1}{l@{\hspace{8pt}}}{Ni--Co} & \multicolumn{1}{l}{$0.56^{+0.05}_{-0.05}$} & & \\
\multicolumn{1}{l@{\hspace{8pt}}}{Co--Ar} & \multicolumn{1}{l}{$0.28^{+0.04}_{-0.04}$} & & \\
\multicolumn{1}{l@{\hspace{8pt}}}{Ni--Ar} & \multicolumn{1}{l}{$0.08^{+0.05}_{-0.05}$} & & \\
\specialrule{\heavyrulewidth}{3pt}{3pt}
\multicolumn{4}{l}{\textit{Directional flux fractions}} \\
\multicolumn{1}{l@{\hspace{8pt}}}{Co flux $< v_{50}$(Ni)} & \multicolumn{1}{l}{$19^{+5}_{-4}$\%} & & \\
\multicolumn{1}{l@{\hspace{8pt}}}{Ar flux $< v_{90}$(Co)} & \multicolumn{1}{l}{$20^{+4}_{-4}$\%} & & \\
\multicolumn{1}{l@{\hspace{8pt}}}{Ar flux $< v_{90}$(Ni)} & \multicolumn{1}{l}{$5^{+5}_{-4}$\%} & & \\
\bottomrule
\end{tabular}
\end{table}

\subsection{Ni, Co, and Ar: A Layered Distribution}

The stable Ni emission lines at [\ion{Ni}{2}]\,6.64\um, [\ion{Ni}{3}]\,7.35 and 11.00\um, and [\ion{Ni}{4}]\,8.41\um\ in \rbs\ at $+$84\,days are well fit by Gaussian distributions with full-width at half-maximum intensity (FWHM) of $\sim$7200\kms\ and kinematic offset with respect to the host-galaxy redshift of $v_{\rm off} \approx -$500\kms. The [\ion{Ni}{4}]\,8.41\um\ line is the most isolated, but the [\ion{Ni}{3}]\,11.00\um\ line has the highest resolution and a clean red side, so we use the [\ion{Ni}{3}]\,11.00\um\ line in \autoref{fig:emissivity}. Notably, this Gaussian-like Ni distribution does not show evidence for excess narrow-core emission, which was observed in the brighter SN\,2022aaiq and 2024gy \citep{Kwok2026}.


The [\ion{Co}{3}]\,11.88\um\ line is essentially isolated, providing a clear probe of the radioactive material in the ejecta. This [\ion{Co}{3}] line in \rbs\ shows clear deviations from the nearly Gaussian structure seen in previous normal SN\,Ia observed with \textit{JWST} \citep{Kwok2023, DerKacy2023, Ashall2024, Kwok2026, Macrie2026}. The line flattens at $v\lesssim 2000$\kms, similar to the low-luminosity SN\,Ia\,2022xkq \citep{DerKacy2024, Kwok2026}. Fitting as a Gaussian with a flattened top, we measure FWHM $\approx 12,000$\kms, $v_{\rm off}\approx-200$\kms, and an inner boundary for the flat top of $v_{\rm inner}\approx2000$\kms.

[\ion{Ar}{3}]\,8.99\um\ cleanly traces the IMEs. As in other normal SN\,Ia, it exhibits a tilted flat-topped profile. In \rbs\ this line is slanted with $\sim10\%$ more flux on the blue side than the red side, about three times more prominent than would be expected from relativistic effects alone \citep[see][their Appendix E]{Blondin2023}. We fit the feature with an asymmetric shell profile in which the outer wings follow a Gaussian and the central region is replaced by a linear interpolation between the fluxes at the inner shell boundaries. We find FWHM $\approx20{,}000$\kms, where the half-maximum is defined relative to the peak flux of the tilted plateau, together with $v_{\rm off}\approx-400$\kms\ and $v_{\rm inner}\approx8000$\kms. The tilt is reproduced by displacing the inner cavity relative to the center of the outer profile, such that the emitting shell extends $\sim350$\kms\ further in velocity space on the blue side than on the red side. The greater effective shell thickness on the blue side produces its higher plateau flux.

To quantify the stratification seen in \autoref{fig:emissivity}, we volume-weight the inverted emissivity profiles, $j_X(v)$, to construct normalized flux-weighted velocity distributions,
\[
p_X(v) \propto j_X(v)\,v^2,
\]
such that $~\int p_X(v)\,{\rm d}v=1~$ and $~p_X(v)\,{\rm d}v~$ gives the fraction of the line flux emitted between $v$ and $v+{\rm d}v$. We define the pairwise overlap coefficient as
\[
\mathrm{OVL}(A,B)=\int \min\left[p_A(v),p_B(v)\right]\,{\rm d}v.
\]
This quantity is bounded between zero and unity, and measures the overlap between the normalized radial flux distributions of the two ions. Uncertainties were propagated using bootstrap realizations of the observed spectrum generated from noise estimated directly from the data near each relevant line, corresponding to S/N $\approx 20$. We report the 16th--84th percentile interval around the bootstrap median. The results are summarized in \autoref{tab:overlap_metrics}.

\autoref{fig:emissivity} and Table~\ref{tab:overlap_metrics} show significant stratification among the stable IGE-, radioactive IGE-, and IME-dominated regions. Half of the [\ion{Ni}{3}] flux is emitted below $v_{50} = 5.0^{+0.4}_{-0.4}\times10^3$\kms, compared with $7.0^{+0.1}_{-0.1}\times10^3$\kms\ for [\ion{Co}{3}] and $11.6^{+0.2}_{-0.1}\times10^3$\kms\ for [\ion{Ar}{3}]. The Ni and Co distributions have $\mathrm{OVL} = 0.56^{+0.05}_{-0.05}$, while only $19^{+5}_{-4}\%$ of the Co flux is emitted interior to the Ni half-flux velocity. These measurements indicate broadly overlapping stable and radioactive IGE regions, but with the radioactive material preferentially concentrated at higher velocities.

The [\ion{Ar}{3}] profile overlaps modestly with the outer tail of the Co distribution ($\mathrm{OVL} = 0.28^{+0.04}_{-0.04}$), with $20^{+4}_{-4}\%$ of the Ar flux interior to the Co 90\%-flux velocity. Its overlap with the Ni distribution is substantially weaker ($\mathrm{OVL} = 0.08^{+0.05}_{-0.05}$), and $5^{+5}_{-4}\%$ of the Ar flux is emitted interior to the Ni 90\%-flux velocity. This is consistent with an IME-dominated outer region containing very little stable Ni.

This separation among these layers is qualitatively more pronounced than in SN\,2024gy, SN\,2022aaiq, and SN\,2021aefx, and is similar---though at higher overall velocities---to the underluminous SN\,2022xkq \citep{Kwok2026}, in which the nucleosynthetic groups are more cleanly separated. Such a layered structure suggests limited macroscopic mixing during the explosion and is broadly characteristic of models containing a detonation, including double detonations and delayed detonations in which the preceding deflagration produces relatively weak mixing. Quantifying nucleosynthetic-layer separations across a larger sample of \snia\ may therefore provide additional constraints capable of distinguishing between explosion mechanisms.

\begin{figure*}
    \centering
    \includegraphics[width=\linewidth]{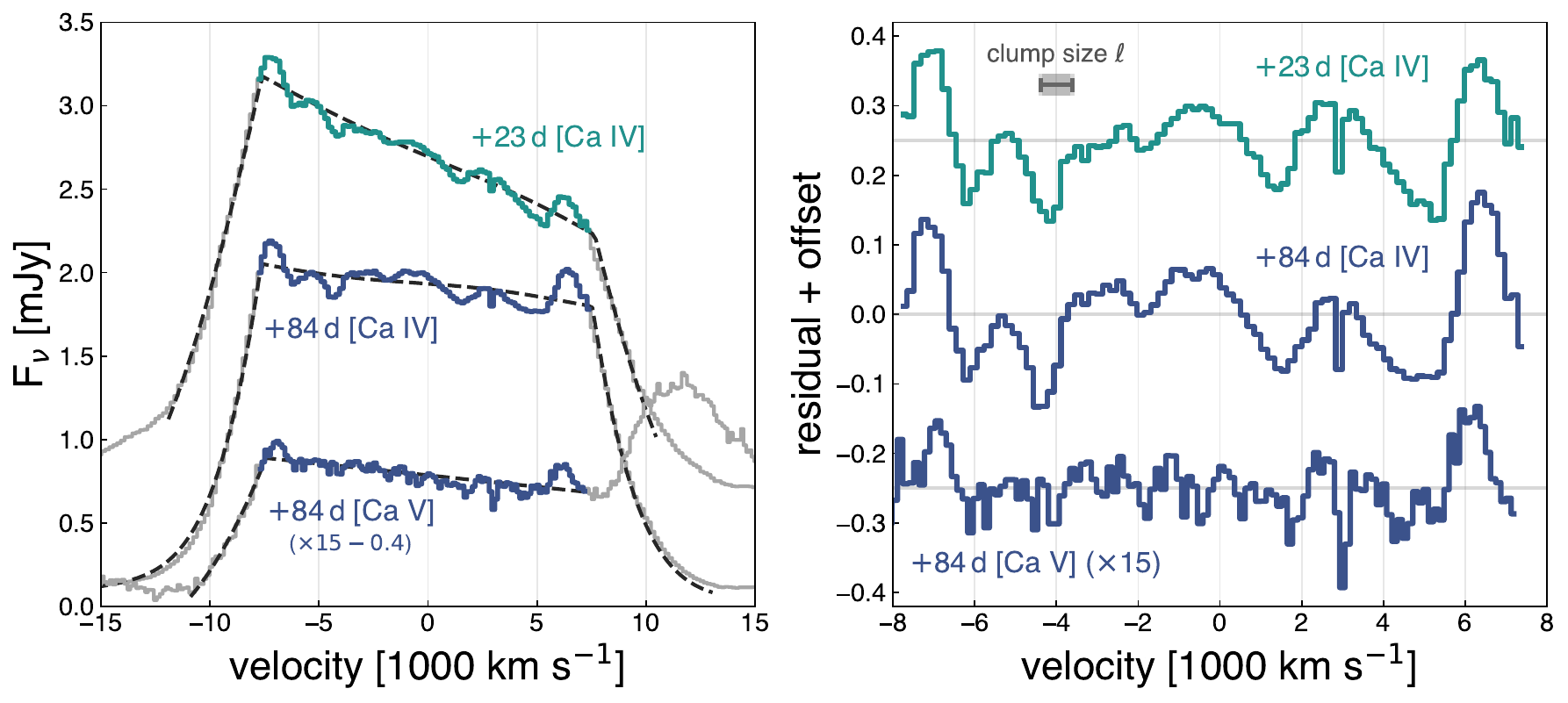}
    \caption{\textit{Left}: [\ion{Ca}{4}]\,3.21\um\ line profiles at $+$23 and $+$84\,days, and [\ion{Ca}{5}]\,4.16\um\ at $+$84\,days, scaled and offset for comparison. The plateau region analyzed for substructure is highlighted in teal ($+$23\,days) and indigo ($+$84\,days). \textit{Right}: Plateau residuals detrended by a high-pass filter. The inferred clump scale $\ell$ is shown as a gray bar.}
    \label{fig:ca_substructure}
\end{figure*}

\subsection{Ca Substructure: Clumping in the Ejecta? \label{sec:ca_substructure}}

The [\ion{Ca}{4}]\,3.21\um\ line exhibits a broad, flat-topped morphology similar to that of [\ion{Ar}{3}]\,8.99\um, but is somewhat narrower at both $+$23 and $+$84\,days. Its FWHM is smaller than that of [\ion{Ar}{3}] by $\sim 1200$\kms\ at $+$23\,days and $2500$\kms\ at $+$84\,days. Thus, the [\ion{Ca}{4}]-emitting material spans a similar velocity range, but is slightly more centrally concentrated than the [\ion{Ar}{3}]-emitting material, broadly consistent with the expected radial ordering of detonation ashes.

Superposed on the bulk profile, the $+$23 and $+$84\,day spectra show prominent fluctuations across the [\ion{Ca}{4}] plateau that repeat between epochs, which we interpret as small-scale structure in the [\ion{Ca}{4}]-emitting distribution (\autoref{fig:ca_substructure}; \textit{left}). Analogous fine structure has been identified in high-S/N observations of very nearby core-collapse SN, including SN\,1985F \citep{Filippenko1989}, SN\,1987A \citep[e.g.,][]{Hanuschik1993, Chugai1994}, SN\,1993J \citep[e.g.,][]{Spyromilio1994, Matheson2000}, and SN\,2011dh \citep{Ergon2015}, where it has been interpreted as evidence of clumping within the ejecta. Such structure can be species dependent; for example, SN\,1993J showed pronounced fluctuations in its O- and Mg-emitting material, whereas its Ca emission lines were relatively smooth \citep{Matheson2000}.

The high S/N ($>$\,100) of the [\ion{Ca}{4}] line allows us to investigate substructure in a thermonuclear SN. At each epoch, we first fit a slanted flat-topped model, following the [\ion{Ar}{3}] fit above and \citet{Kwok2026}, and isolate the plateau between $-$7800 and $+$7500\kms\ by subtracting the fit. We then apply a Gaussian high-pass filter by subtracting a smoothed version of the residual profile using a 15-pixel ($\sim$2500\kms) smoothing window. This detrending suppresses remaining broad variations on scales of a few $\times10^3$\kms\ and larger while preserving the smaller-scale fluctuations.

This procedure is closely related to the approach used for SN\,1993J by \cite{Matheson2000}, where a boxcar-smoothed line profile was subtracted to isolate small-scale fluctuations. In \rbs, the well-defined plateau allows us to remove the bulk profile with a physically motivated fit before detrending, reducing sensitivity to the adopted smoothing scale and to edge artifacts. As a cross-check, residuals obtained using a $\sim3700$\kms\ boxcar smoothing scale are strongly correlated with those from our adopted method ($r\approx0.9$), indicating that both recover essentially the same fluctuations.

The $+$23 and $+$84\,day [\ion{Ca}{4}] residuals are themselves strongly correlated ($r\approx 0.92$), showing that the same pattern persists at fixed velocity between the two epochs (\autoref{fig:ca_substructure}). This is inconsistent with independently realized random noise. SN\,2022aaiq and SN\,2024gy, observed at high S/N with the same NIRSpec configuration, also exhibit structure across the [\ion{Ca}{4}]\,3.21\um\ plateau \citep{Kwok2026}, with different fluctuation patterns. This object-dependent morphology precludes a fixed-pattern instrumental origin. In \rbs, the persistence of the pattern as the underlying continuum fades between $+$23 and $+$84\,days further disfavors contamination by another spectral component present only at $+$23\,days.

To characterize the velocity scale of the substructure, we calculate the autocorrelation function (ACF) of the detrended residuals. The ACF measures the similarity between the residual profile and a copy shifted by a given velocity lag. We define the observed correlation length, $L_{1/e}$, as the lag at which this similarity falls to $1/e$ of its zero-lag value. It therefore measures the characteristic velocity width of the fluctuations, rather than the physical size of an individual clump directly.

We measure correlation lengths of $L_{1/e}=530^{+60}_{-60}$\kms\ at $+$23\,days and $L_{1/e}=590^{+90}_{-70}$\kms\ at $+$84\,days, with uncertainties estimated by block-bootstrap resampling of the residuals. The agreement between epochs indicates that the characteristic fluctuation width remains approximately constant during homologous expansion. Because these values are less than twice the instrumental resolution of $\sim300$\kms, we test whether instrumental broadening could transform random noise into features of the observed width. We generate white-noise profiles, broaden them to the NIRSpec resolution, and process them in the same way as the data. This produces a characteristic correlation length of only $240^{+30}_{-30}$\kms. The observed correlation lengths are therefore substantially larger than expected from instrumental broadening of random noise, confirming that they trace resolved structure in the line profile.

We infer the intrinsic clump scale using a forward model based on the statistical clump framework of \cite{Chugai1994}, generalized to a spherical shell. We populate the shell, extending from $v_{\rm in}=7000$\kms\ to $v_{\rm out}=15{,}000$\kms, with randomly distributed clumps and calculate the resulting projected line profiles. Each clump is represented by a Gaussian emissivity profile with intrinsic FWHM $\ell$. The simulated profiles are then broadened to the NIRSpec resolution, sampled onto the observed velocity grid, and processed with the same detrending and ACF measurement as the data. This provides a direct calibration between the intrinsic clump scale $\ell$ and the measured $L_{1/e}$. 

We fit a single characteristic clump scale jointly to the ACF measurements from both epochs and obtain $\ell=780^{+100}_{-100}$\kms (\autoref{fig:ca_substructure}, \textit{right}). Tests with simulated profiles of known input size show that clumps near this scale remain distinguishable after instrumental broadening; only clumps $\lesssim200$\kms\ become unresolved. Here, $\ell$ is the FWHM of the Gaussian emissivity profile assigned to each clump and is interpreted below as a diameter-equivalent characteristic scale.

The fractional root-mean-square (RSM) amplitude provides complementary information about the strength of the fluctuations and, together with the clump scale $\ell$, constrains the volume filling factor \citep{Chugai1994, Spyromilio1994}. We define
\[
\epsilon \equiv
\frac{\left\langle \delta F_{\rm hp}(v)^2\right\rangle^{1/2}}
{\overline{F}_{\rm Ca}},
\]
where $\delta F_{\rm hp}(v)$ is the mean-subtracted, high-pass-filtered residual flux density and $\overline{F}_{\rm Ca}$ is the mean continuum-subtracted fitted [\ion{Ca}{4}] flux density across the plateau. For the $+$23\,day spectrum, we subtract an estimated flat 0.9\,mJy underlying continuum when calculating $\overline{F}_{\rm Ca}$; no continuum correction is applied at $+$84\,days. 

We measure $\epsilon=0.032\pm0.008$ at $+$23\,days and $\epsilon=0.036\pm0.009$ at $+$84\,days. The uncertainties include both measurement precision, estimated through block-bootstrap resampling of the observed residuals, and the realization-to-realization scatter expected from the finite number and random placement of clumps, estimated with Monte Carlo simulations. Although the fractional RMS amplitudes are only 3--4\%, individual peak-to-peak excursions reach  $\sim10$\% across the interior of the plateau. The excursions increase to $\sim16$\% when the plateau boundaries are included; however, these regions are also more sensitive to uncertainties in the fitted bulk profile and the high-pass filtering. We therefore use $\epsilon$ as the primary statistic describing the substructure amplitude.

To estimate the volume filling factor $f$, we use Equation~11 of \cite{Chugai1994}, generalized from a filled sphere to our adopted thick-shell geometry. In this calculation, we use $\ell$ as the diameter of a spherical clump, so that its radius is $u=\ell/2$. The ACF calibration and amplitude calculation therefore use different idealized clump profiles, Gaussian emissivity elements for the projected line shape and uniform spheres for the filling-factor calculation, but share the same diameter-equivalent size convention.

Combining the ACF and fractional RMS constraints from both epochs gives $f=0.10^{+0.01}_{-0.01}$. Together with $\ell=780^{+100}_{-100}$\kms, this corresponds to $N_c=5100^{+2200}_{-1400}$ Ca-emitting clumps for the adopted shell and clump geometry. The characteristic clump scale $\ell$ is constrained primarily by the widths of the fluctuations and is only weakly sensitive to the adopted shell boundaries. In contrast, both $f$ and $N_c$ depend strongly on the assumed emitting volume and are therefore sensitive to the adopted $v_{\rm in}$ and $v_{\rm out}$. We consequently regard $\ell$ as the more robust inference, while $f$ and $N_c$ should be interpreted as geometry-dependent estimates. Combined with the strong correlation between the residual profiles ($r=0.92$), the agreement of the ACF correlation lengths and fractional RMS amplitudes indicates that the data are consistent between epochs with the same characteristic clump population remaining fixed in velocity space ($\chi^2/{\rm dof}=0.30$ for two degrees of freedom (dof)).

We find additional evidence for Ca-associated substructure in the much weaker [\ion{Ca}{5}]\,4.16\um\ line at $+$84\,days (\autoref{fig:ca_substructure}). Over their common velocity interval of $-7700$ to $+7200$\kms, the detrended [\ion{Ca}{4}] and [\ion{Ca}{5}] residuals are correlated at $r=0.66$. Notably, the narrow dip in the residuals near $\sim3000$\kms\ appears to be present in [\ion{Ca}{5}] as well, possibly suggesting a physical origin rather than a bad-pixel artifact. Because [\ion{Ca}{5}] has substantially lower S/N, measurement noise is expected to reduce the observed correlation. The correspondence between two independent Ca transitions further supports a physical origin for the fluctuations in Ca emission in the ejecta.

The line-profile fluctuations trace variations in emissivity, however, and do not necessarily imply an equally inhomogeneous Ca abundance distribution. Producing Ca$^{3+}$ and Ca$^{4+}$ requires ionization energies of 50.9 and 67.3\,eV, respectively. By comparison, the thresholds for producing Ar$^{+}$ and Ar$^{2+}$ are 15.8 and 27.6\,eV, while those for producing the doubly ionized IGE species Fe$^{2+}$, Co$^{2+}$, and Ni$^{2+}$ are only $\sim16$--18\,eV \citep{NIST_ASD}.\footnote{\href{https://www.nist.gov/pml/atomic-spectra-database}{NIST Atomic Spectra Database}} [\ion{Ca}{4}] and [\ion{Ca}{5}] may therefore be especially sensitive to spatial variations in the local ionization conditions.

The similar structure observed in [\ion{Ca}{4}] and [\ion{Ca}{5}], together with the smoother lower-ionization Ar and IGE profiles, might reflect a patchy ionization state rather than, or in addition to, a clumpy Ca abundance distribution. Because we do not have a sufficiently clean line from a lower Ca ionization stage, we cannot  uniquely distinguish between compositional clumping and spatial variations in the ionization balance.

\begin{figure*}[tb]
    \centering
    \includegraphics[width=\linewidth]{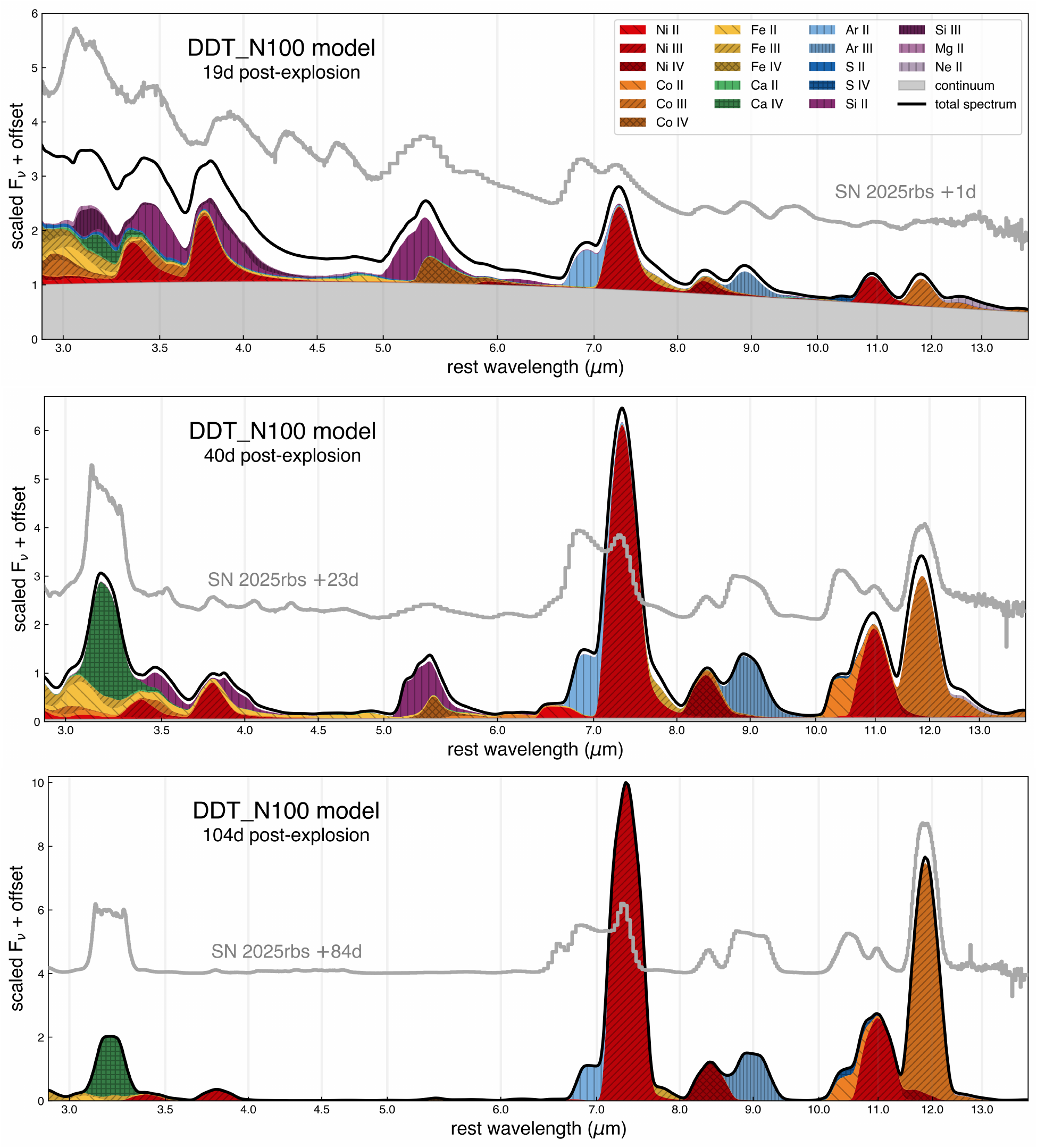}
    \caption{Illustrative ion decomposition of the \texttt{DDT\_N100} model, used to guide line identifications. No model parameters have been optimized to reproduce the observed line strengths, widths, or ionization states. Model spectra at 18.73, 40.15, and 104.1\,days post-explosion, scaled to 14.5\,Mpc, are compared to \rbs\ at $+$1, $+$23, and $+$84\,days relative to \textit{B}$_{\rm max}$ (gray; assuming an $\sim$18\,day rise). The \rbs\ spectra are vertically offset for clarity. Contributions from individual ions are indicated by colored, hatched regions, with the continuum shown in gray. The differential single-ion components do not sum exactly to the total model spectrum, particularly at early epochs, because removing one ion alters the radiative response of the remaining ions.}
    \label{fig:models}
\end{figure*}

\begin{figure*}[tb]
    \centering
    \includegraphics[width=\linewidth]{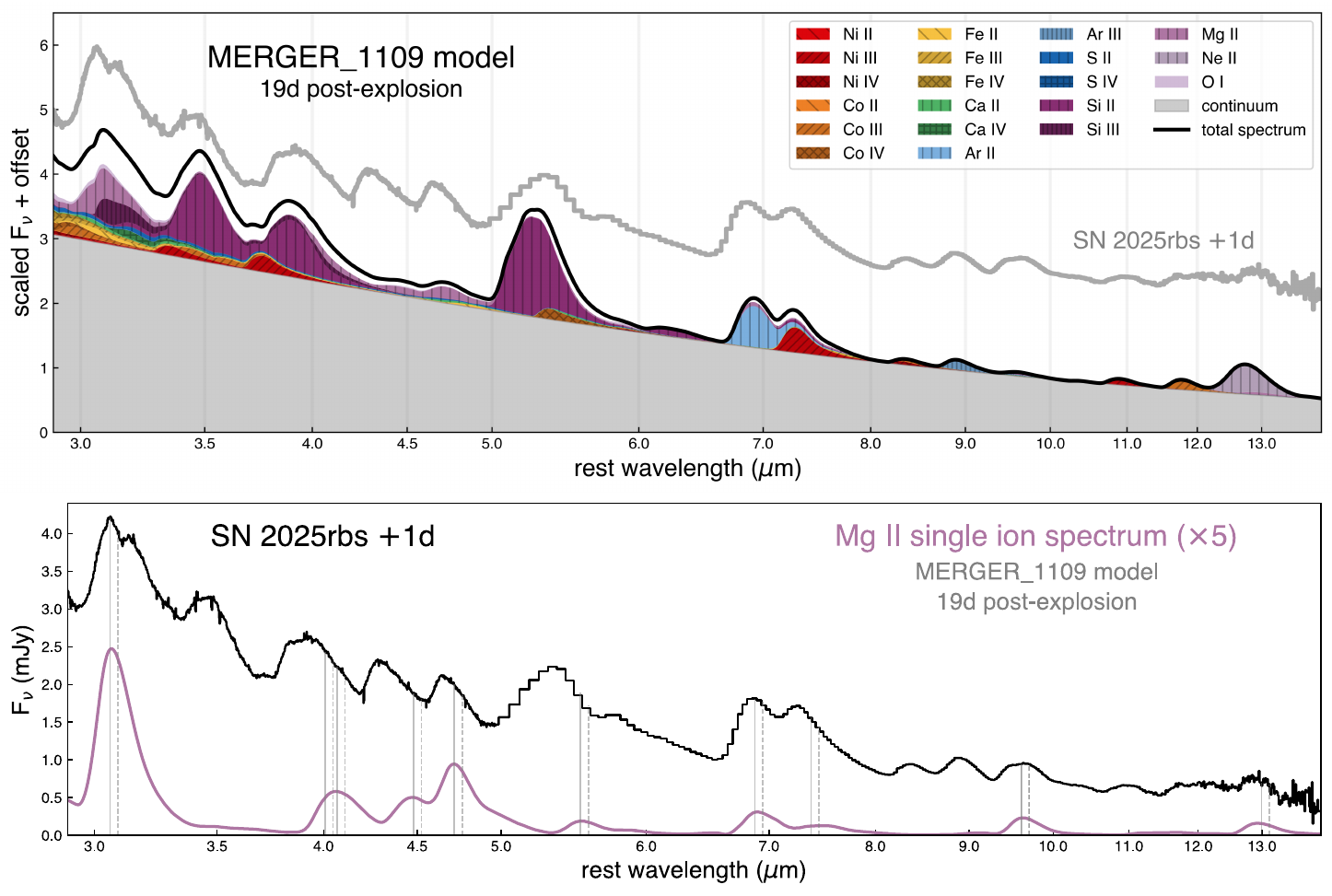}
    \caption{\textit{Top}: Illustrative ion decomposition of the \texttt{MERGER\_1109} model at $\sim$20\,days post-explosion, scaled to 14.5\,Mpc. We use the comparison to investigate possible \ion{Mg}{2} identifications rather than to propose a violent merger origin for \rbs. The model produces stronger \ion{Mg}{2} contributions than \texttt{DDT\_N100}, but still underpredicts several candidate \ion{Mg}{2} features. It also predicts prominent [\ion{Ne}{2}]\,12.82\um\ emission whose velocity profile is inconsistent with the observed feature near 13\um. \textit{Bottom}: Differential single-ion \ion{Mg}{2} spectrum from the \texttt{MERGER\_1109} model compared to \rbs\ at $+$1\,day. The \ion{Mg}{2} spectrum is multiplied by a factor of five to highlight the correspondence between predicted transitions and observed features near 4.6, 9.6, and 13.0\um; the required scaling also demonstrates that these features are substantially stronger than predicted. The strongest model \ion{Mg}{2} lines in the range 3--14\um\ are marked at their rest wavelengths (dashed gray) and at $-3000$\kms\ (solid gray), corresponding to the model peak velocities.}
    \label{fig:Mg_merger}
\end{figure*}

\section{Radiative-Transfer Models \label{sec:models}}

The complex combination of permitted lines, forbidden lines, and continuum emission in the $+$1 and $+$23\,day MIR spectra requires radiative-transfer calculations to identify the ions contributing to individual features. We use the models primarily as qualitative line-identification and interpretation tools; no model parameters are optimized to reproduce the observed line strengths, widths, or ionization states. We compute model spectra using the CMFGEN code of \cite{Hillier2012}, post-processing selected time steps of a time-dependent calculation to include the latest atomic data used by \citet{Blondin2023}.

\autoref{fig:models} shows the \texttt{DDT\_N100} calculation, using the spherically averaged ejecta structure of the delayed-detonation N100 model of \cite{Seitenzahl2013} as inputs, at 18.73, 40.15, and 104.1\,days post-explosion. These epochs correspond closely to the phases of the \rbs\ \textit{JWST} observations at $+$1, $+$23, and $+$84\,days relative to \textit{B}$_{\rm max}$ (18\,day rise time). To isolate the contribution of a given ion, we compute an observer-frame spectrum with that ion excluded and subtract it from the calculation containing the full atomic set. We plot these differential single-ion spectra cumulatively to show the extent of line blending and identify the dominant ions contributing to individual features. The components do not sum exactly to the total spectrum because removing one ion changes the response of the remaining ions to the radiation field. This nonadditivity is strongest at the earliest, still optically thick epoch, and becomes less important as the ejecta become optically thin. The early-time ion decompositions should therefore be interpreted as guides to the dominant line contributors.

The \texttt{DDT\_N100} calculation indicates that \ion{Si}{2} and \ion{Si}{3} contribute substantially to several features between 3\um\ and 6\um\ that fade between $+$1 and $+$23\,days. Forbidden [\ion{Ni}{3}] and [\ion{Co}{4}] emission also contribute prominently in this wavelength range and fade from $\sim$20 to 100\,days post-explosion, although more gradually than the permitted Si features. Many of the dominant forbidden transitions between 3 and 14\um\ are already present in the earliest model and can be associated with features in the observed $+$1\,day spectrum. The main exceptions are [\ion{Ni}{2}] and [\ion{Co}{2}], which strengthen at later epochs as the ejecta recombine to lower ionization stages.

Although the calculation successfully identifies many of the ions contributing to the spectrum, it does not reproduce all of the observed lines or their relative strengths. At 104.1\,days, the model is too highly ionized, most clearly demonstrated by the overly strong [\ion{Ni}{3}] and [\ion{Co}{3}] emission. Similar late-time overionization has also been found in other CMFGEN and ARTIS calculations \citep{Shingles2022, Blondin2023, Pollin2025, Pollin2026}. The velocity distributions also differ from the observations: the modeled Ni features are broader than observed, whereas the modeled Ar and Ca features are narrower. Thus, while the calculation is useful for identifying the dominant emitting ions, neither its ionization balance nor its chemical distribution in velocity space reproduces the data in detail. Such discrepancies are not unexpected, as the model was not optimized to match \rbs.

The $+$1\,day spectrum also contains several features that are absent or substantially weaker in the 18.73\,day \texttt{DDT\_N100} calculation, particularly near 4.3, 4.6, 9.6, and 13.0\um. The 4.3\um\ feature can be partially or wholly attributed to a permitted \ion{Ni}{2} transition, although the line would need to be stronger than predicted. The remaining features have few plausible identifications in either the model line list or the Atomic Line List \citep{vanHoof2018}. We find that the most compelling candidate is \ion{Mg}{2}, which has multiple transitions capable of contributing across this wavelength range.

To investigate the weak predicted \ion{Mg}{2} emission, we compare the data with the \texttt{MERGER\_1109} CMFGEN calculation, based on the spherically averaged ejecta structure of the violent WD--WD merger model of \cite{Pakmor2012}, using the same differential single-ion approach. Violent-merger ejecta retain larger abundances of unburned and carbon-burning products, including C, O, Ne, and Mg, at relatively low velocities. Peculiar thermonuclear SN with late-time [\ion{O}{1}] or [\ion{Ne}{2}] emission have therefore often been interpreted in the context of violent mergers \citep[e.g.,][]{Kromer2013, Dimitriadis2023, Blondin2023, Siebert2023, Siebert2024, Kwok2024, Pakmor2026}. Here, however, we use \texttt{MERGER\_1109} primarily to test how the different Mg abundance and ionization structure affect the candidate \ion{Mg}{2} features, rather than as a proposed explosion model for \rbs.

As shown in \autoref{fig:Mg_merger}, the \texttt{MERGER\_1109} model produces stronger \ion{Mg}{2} emission than the \texttt{DDT\_N100} model. Comparison of the Mg abundance, ionization fractions, and the line-forming region of the 4.76\um\ transition in both models shows that the difference is a combination of composition and ionization. The \texttt{MERGER\_1109} model contains more Mg at low velocities and maintains a larger fraction of it as \ion{Mg}{2}. Its lower ionization is likely related in part to the higher ejecta densities \citep{Blondin2023}. Consequently, the \texttt{MERGER\_1109} model produces additional flux near the observed 4.6, 9.6, and 13.0\um\ features and better reproduces several of their characteristic wavelengths and velocities. Nevertheless, the predicted MIR \ion{Mg}{2} emission remains substantially weaker than observed.

The \texttt{MERGER\_1109} calculation also improves the agreement with several other observed features, including the \ion{Si}{2} features near 3.5 and 4.0\um\ and the relative strengths of [\ion{Ar}{2}] and [\ion{Ni}{3}]. This may reflect its lower overall ionization, its different distribution of elements in velocity space, or both. The model does not, however, provide a complete explanation of the spectrum. In particular, it predicts strong [\ion{Ne}{2}]\,12.82\um\ emission whose velocity profile is inconsistent with the observed feature near 13.0\um: identifying this feature as [\ion{Ne}{2}] would require a redshift of several thousand \kms, whereas the other lines at this epoch are blueshifted. At nebular epochs, the model [\ion{Ne}{2}] line becomes even stronger, while the corresponding observed feature disappears, placing the model in direct conflict with the later epochs of \rbs.

The combination of stronger-than-predicted candidate \ion{Mg}{2} emission and weak or absent [\ion{Ne}{2}] is unlikely to result simply from an abundance distribution rich in Mg but deficient in Ne, since the two elements are produced in neighboring burning regimes. Instead, the discrepancy may reflect the ionization and excitation balance in the calculations. Mg and Ne also probe different ionization thresholds: \ion{Mg}{2} is ionized to \ion{Mg}{3} at 15.0\,eV, while ionization of neutral Ne to \ion{Ne}{2} requires 21.6\,eV. Changes in the ionization state can therefore alter the populations responsible for the Mg and Ne emission differently. Moreover, as discussed earlier, the candidate MIR \ion{Mg}{2} transitions appear to be populated primarily through recombination cascades and therefore depend on the flow from \ion{Mg}{3} into excited \ion{Mg}{2} states, rather than simply on the total \ion{Mg}{2} population. The weakness of the modeled MIR \ion{Mg}{2} features should therefore not be interpreted as a straightforward global deficit of \ion{Mg}{2}.

In the bottom panel of \autoref{fig:Mg_merger}, we compare the differential single-ion \ion{Mg}{2} spectrum from \texttt{MERGER\_1109} directly with the $+$1\,day MIR spectrum. We multiply the model's \ion{Mg}{2} contribution by a factor of five solely to emphasize the correspondence between the predicted transitions and the observed emission near 4.6, 9.6, and 13.0\um. These three otherwise unexplained features align closely with \ion{Mg}{2} transitions, while other strong \ion{Mg}{2} lines between 3 and 14\um\ are also consistent with contributing to weaker or blended structures near 3.0, 4.1, 4.5, 5.6, 7.0, and 7.5\um. We therefore tentatively identify the 4.6, 9.6, and 13.0\um\ features as predominantly \ion{Mg}{2} emission.

However, this scaling should not be interpreted as a physically consistent global rescaling of the Mg abundance or ion population. We compare the permitted \ion{Mg}{2}\,1.0927\um\ feature in the models to the $+$1\,day spectrum and find that in the unscaled \texttt{MERGER\_1109} calculation, the \ion{Mg}{2} contribution near 1.0927\um\ is already comparable to, and slightly stronger than, the observed feature, whereas the \texttt{DDT\_N100} model predicts a somewhat weaker contribution. Scaling the merger \ion{Mg}{2} single-ion spectrum by the factor used to match the MIR features thus dramatically overpredicts the 1.0927\um\ line. Unlike the MIR transitions, the lower level of the 1.0927\um\ line is expected to be populated primarily by photoexcitation through strong UV \ion{Mg}{2} transitions, making its strength more directly sensitive to the \ion{Mg}{2} population and radiation field. Thus, the disagreement is transition dependent rather than a uniform deficit in all \ion{Mg}{2} emission. Because the NIR photoexcited line and MIR recombination-cascade lines respond differently to the \ion{Mg}{2}/\ion{Mg}{3} ionization balance, their simultaneous observation provides a potentially sensitive panchromatic diagnostic of the Mg ionization and excitation states.

\section{Discussion \& Conclusions}

We present the earliest MIR spectroscopic sequence of a normal \snia\ to date, including the first maximum-light MIR spectrum, obtained with \textit{JWST} for \rbs\ at $+$1, $+$23, and $+$84\,days post \textit{B}-band maximum. The MIR spectra evolve rapidly: at $+$1\,day the spectrum exhibits a continuum superposed with a mix of permitted and forbidden features; by $+$23\,days the continuum has faded significantly and the spectrum is dominated by growing forbidden emission lines; and by $+$84\,days the spectrum is fully nebular, with negligible continuum and fully developed forbidden-line profiles. Our main findings are as follows.

\begin{itemize}
    \item The early MIR spectral sequence of \rbs\ directly traces the transition from the photospheric to nebular phases. At maximum light, forbidden [\ion{Ni}{3}], [\ion{Ni}{4}], [\ion{Ar}{2}], and [\ion{Ar}{3}] emission is already emerging in the MIR despite the presence of a continuum and permitted-line emission. The [\ion{Ar}{3}]\,8.99\um\ profile subsequently evolves from optically thick at $+$1\,day to a fully nebular flat-topped profile with sharp blue and red edges by $+$84\,days, when the MIR continuum has disappeared. Over the same interval, the MIR contribution to the 0.4--14\um\ flux grows from $<1\%$ to $\sim7\%$.
    \item The nebular MIR spectrum reveals a strongly stratified distribution of nucleosynthetic products, with stable Ni concentrated at the lowest velocities, radioactive Co at intermediate velocities, and Ar occupying an outer shell. Half of the [\ion{Ni}{3}], [\ion{Co}{3}], and [\ion{Ar}{3}] flux is emitted below $\sim5000$, 7000, and $11,500$\kms, respectively. The Ni and Co distributions overlap substantially ($\mathrm{OVL}\approx0.56$), whereas the overlap of Ar with Co ($\mathrm{OVL}\approx0.28$) and particularly Ni ($\mathrm{OVL}\approx0.08$) is much smaller. The flat-topped [\ion{Co}{3}]\,11.88\um\ profile further indicates a deficit of radioactive emission within $\sim2000$\kms.
    \item The [\ion{Ca}{4}]\,3.21\um\ profile shows persistent small-scale substructure at both $+$23 and $+$84\,days, with strongly correlated residual profiles ($r=0.92$), fractional RMS amplitudes of $\sim3$--4\%, and a characteristic clump scale of $\ell\approx800$\kms. For an adopted thick-shell geometry, the amplitude and scale correspond to a volume filling factor of $f\approx0.10$ and $\sim5100$ emitting clumps, although these latter quantities depend strongly on the assumed emitting volume. Similar fluctuations in [\ion{Ca}{5}]\,4.16\um\ support a physical origin associated with the Ca-emitting material. The observed emissivity fluctuations of these highly ionized Ca lines could reflect compositional clumping, spatial variations in ionization, or both.
    \item Radiative-transfer calculations identify several otherwise unexplained features at 4.6, 9.6, and 13.0\um\ in the $+$1\,day spectrum as candidate \ion{Mg}{2} transitions. The \texttt{MERGER\_1109} model produces stronger MIR \ion{Mg}{2} emission than \texttt{DDT\_N100} through a combination of enhanced Mg abundance and different ionization conditions, but both calculations substantially underpredict the observed MIR features. However, they do approximately reproduce the NIR \ion{Mg}{2}\,1.0927\um\ feature. The NIR line is likely populated primarily through photoexcitation, whereas the high-lying MIR transitions are probably populated through recombination cascades.
\end{itemize}

The smooth Gaussian-like Ni profiles of \rbs\ distinguish it from several brighter normal \snia\ previously observed with \textit{JWST}, which show either narrow-core Ni components, broken-slope structure, or both \citep[e.g., SN\,2022aaiq, SN\,2024gy, and tentatively SN\,2021aefx;][]{Kwok2026}. In this respect, \rbs\ more closely resembles SN\,2023qov \citep{Macrie2026}, which similarly shows smooth, Gaussian-like Ni profiles. They are also remarkably similar photometrically, with essentially identical decline rates of $\Delta m_{15}(B)=1.47$\,mag and comparable peak luminosities, although \rbs\ is 0.1--0.2\,mag brighter. \rbs\ and SN\,2023qov have narrower Ni distributions compared with the brighter events, but broader and stronger Ni emission than in the subluminous SN\,2022xkq \citep{DerKacy2024, Ashall2024, Kwok2026}, suggesting that the morphology and extent of stable IGE material vary systematically across normal \snia.

The radial separation of the stable IGE-, radioactive IGE-, and IME-dominated regions indicates limited macroscopic mixing, but its interpretation depends on the explosion scenario. In a near-$M_{\rm Ch}$ delayed detonation, stable neutron-rich Ni is produced primarily during the high-density deflagration, whereas much of the radioactive $^{56}$Ni is synthesized later in the detonation; a separation between the Ni- and Co-emitting regions could therefore reflect the distinct spatial distributions of deflagration and detonation ashes. A relatively weak deflagration would generally produce less stable Ni and less pre-expansion, allowing the subsequent detonation to occur at higher density and synthesize more $^{56}$Ni. Such models are therefore typically associated with brighter explosions, complicating a simple interpretation of the strong stratification in \rbs\ as evidence for a weak deflagration alone. Alternatively, a sub-$M_{\rm Ch}$ detonation can naturally produce strongly layered ejecta without a preceding deflagration, although its stable Ni originates primarily by the progenitor neutron excess likely inherited through metallicty and potential $^{22}$Ne settling. Measurements of nucleosynthetic-layer overlap across a larger sample may offer an independent axis for distinguishing among explosion scenarios.

The Ca-associated substructure provides a different probe of the explosion on much smaller velocity scales. Because the fluctuations are seen in more than one highly ionized Ca transition but are much weaker or absent in the lower-ionization Ar and IGE profiles, their origin is unlikely to be a simple global clumping of the ejecta. Instead, they may trace inhomogeneities specific to the Ca-emitting regions, spatial variations in the ionization state, or a combination of composition and ionization structure rather than abundance structure directly. 

A natural interpretation of small-scale structure in thermonuclear SN ejecta is that it reflects plumes and inhomogeneities generated during the deflagration phase, where Rayleigh--Taylor instabilities can produce substantial mixing and large-scale ejecta inhomogeneities. However, deflagration is not the only mechanism capable of producing small-scale fluctuations. Three-dimensional double-detonation calculations provide one alternative context: \cite{Pollin2025} find viewing-angle-dependent line-profile structure even in the absence of a deflagration. In those calculations, however, comparable structure generally appears across multiple species, unlike in \rbs\ where the substructure is only clearly detected in [\ion{Ca}{4}] and [\ion{Ca}{5}]. The difference may reflect the sensitivity of these Ca lines to spatial variations in the ionization state, which may not be reproduced by the models, or because the fluctuations have a different physical origin altogether.

The early-time \ion{Mg}{2} emission provides a complementary probe of both the outer carbon-burning layers and the ionization physics of the ejecta. In particular, the contrasting behavior of the NIR and MIR \ion{Mg}{2} transitions provides a useful constraint on the radiative-transfer calculations. Because the NIR feature is likely dominated by photoexcitation while the high-lying MIR transitions are populated primarily through recombination cascades, their relative strengths probe different aspects of the Mg ionization and excitation balance. The inability of a simple rescaling of the Mg contribution to reproduce both wavelength regimes argues against interpreting the MIR discrepancy as merely an underestimate of the Mg abundance. More generally, species with measurable transitions spanning multiple wavelength regimes provide strong tests of radiative-transfer calculations because a single ionization and excitation structure must reproduce them simultaneously. \ion{Mg}{2} is a particularly valuable early-time example, since its NIR and MIR transitions probe different excitation pathways and the rapid disappearance of the MIR features requires sufficiently early MIR spectroscopy.

The velocity structure of the candidate \ion{Mg}{2}\,9.71\um\ feature offers a complementary constraint on the radial distribution of carbon-burning products. Its extended blue wing may indicate that Mg occupies layers exterior to the Ar-emitting region, as expected in stratified explosions (e.g., delayed- and double-detonation scenarios) in which carbon burning occurs at lower densities than the production of heavier IMEs, although the observed profile also depends on excitation and optical-depth effects. In sub-$M_{\rm Ch}$ double detonations, the velocity extent of the Mg-rich region is expected to depend on primary WD mass, given that more-massive primaries have steeper density profiles and less material at the relatively low densities that produce Mg, leading to a narrower Mg distribution, potentially making sufficiently early MIR spectroscopy sensitive to both the explosion mechanism and progenitor structure. 

The observations presented here highlight both the promise and the current limitations of radiative-transfer modeling in the transitional phase. The coexistence of permitted and forbidden lines at $+$1\,day, the early emergence of forbidden emission, and the underprediction of MIR \ion{Mg}{2} strength all point to the need for fully time-dependent models capable of self-consistently treating the partially optically thick ejecta. Such models are computationally expensive and challenging, but the new constraints from early-time MIR spectroscopy offer additional motivation for their development. Reproducing the full panchromatic spectral sequence of \rbs\ across the photospheric, transitional, and nebular phases will define a demanding new benchmark for the next generation of \snia\ radiative-transfer models.

The panchromatic spectral sequence of \rbs\ demonstrates the value of extending observations of normal \snia\ into the MIR during early and transitional phases. Because optical depth is strongly wavelength dependent, MIR spectroscopy can access ejecta layers and transitions that remain obscured or blended at shorter wavelengths, providing information complementary to contemporaneous optical and NIR observations. In this work, we focus primarily on the MIR evolution and the constraints it provides on the ejecta structure. The broader panchromatic sequence offers additional opportunities to connect these MIR diagnostics with the contemporaneous optical and NIR spectra, which we will explore in future work. Together with improved radiative-transfer calculations, such panchromatic observations can provide increasingly stringent tests of \snia\ explosion models.

\begin{acknowledgements}

This work is based on data obtained with the NASA/ESA/CSA \textit{JWST} as part of program \#09255. We thank Milo Docher for their consistently excellent work scheduling the \textit{JWST} observations, Nimisha Kumari for assistance with the NIRSpec observations, and Andreea Petric for help with the MIRI observations. The data were obtained from the Mikulski Archive for Space Telescopes at the Space Telescope Science Institute (STScI), which is operated by the Association of Universities for Research in Astronomy (AURA), Inc., under National Aeronautics and Space Administration (NASA) contract NAS 5-03127 for \textit{JWST}. Support for this program at Northwestern University was provided by NASA through grant JWST-GO-09255.001.
L.A.K. is supported by NASA through an NHFP Hubble Fellowship grant HF2-51579.001-A awarded by STScI.

F.P.C.\ acknowledges funding from STFC grant ST/X00094X/1.
A.F.\ acknowledges support by the European Research Council (ERC) under the European Union's Horizon 2020 research and innovation program (ERC Advanced Grant KILONOVA No.\ 885281).
J.T.H.\ acknowledges support from NASA through the NASA Hubble Fellowship grant HST-HF2-51577.001-A, awarded by STScI. STScI is operated by AURA, Inc., 
under NASA contract NAS5-26555.
W.B.H.\ acknowledges support from the National Science Foundation (NSF) Graduate Research Fellowship Program under grant 2236415.
Any opinions, findings, and conclusions or recommendations expressed in this material are those of the author(s) and do not necessarily reflect the views of the NSF.

This work makes use of data from the Las Cumbres Observatory network of robotic telescopes. The LCO team is supported by NSF grant AST-2308113.
C.L.\ and A.A.M.\ are supported by DoE award \#\,DE-SC0025599, while A.A.M.\ is also supported by Cottrell Scholar Award \#\,CS-CSA-2025-059 from Research Corporation for Science Advancement.
K.~Maeda\ acknowledges support from JSPS KAKENHI grants (JP24KK0070, JP24H01810, and 23H04894).
K.~Maguire\ acknowledges funding from Horizon Europe ERC grant 101125877.
K.~Medler\ acknowledges support from NASA/JWST grants GO-02114, GO-02122, GO-03726, GO-04217, GO-04436, GO-04522, GO-05057, GO-05290, GO-06023, GO-06213, GO-06583, and GO-06677. Support for these programs was provided by NASA through a grants from STScI, which is operated by AURA, Inc., under NASA contract NAS5-03127.

N.R.\ is supported by a Northwestern University Presidential Fellowship Award. Zwicky Transient Facility access for N.R.\ was supported by Northwestern University and the Center for Interdisciplinary Exploration and Research in Astrophysics (CIERA).
Time-domain research by the University of Arizona team and D.J.S.\ is supported by NSF grants 2308181, 2407566, and 2432036.
M.~Shrestha\ acknowledges funding from the Australian Research Council (ARC) Centre of Excellence CE230100016.
M.~Singh\ acknowledges financial support provided under the National Post Doctoral Fellowship (N-PDF; File Number PDF/2023/002244) by the Science \& Engineering Research Board (SERB), Anusandhan National Research Foundation (ANRF), Government of India.
T.S.\ acknowledges Hungarian NKFIH OTKA grant K-142534.
T.T.\ acknowledges support from NASA/JWST grant GO-02072.011, NSF grant AST-2205314, and NASA/ADAP award 80NSSC23K1130.
J.H.T.\ acknowledges Horizon Europe ERC grant 101125877.
S.V.\ and the UC Davis time-domain research team acknowledge support from NSF grant AST-2407565.
J.V.\ is supported by Hungarian NKFIH OTKA Grant No.\ K-142534.
A.V.F.’s research group at UC Berkeley acknowledges financial assistance from the Christopher R. Redlich Fund, Gary and Cynthia Bengier, Clark and Sharon Winslow, Alan Eustace and Kathy Kwan (W.Z. is a Bengier-Winslow-Eustace Specialist in Astronomy), Timothy and Melissa Draper, Briggs and Kathleen Wood, Ellyn and Alan
Seelenfreund (T.G.B. is Draper-Wood-Seelenfreund Specialist in Astronomy), and numerous other donors.   

A major upgrade of the Kast spectrograph on the Shane 3\,m telescope at Lick Observatory, led by Brad Holden, was made        possible through gifts from the Heising-Simons Foundation, William and Marina Kast, and the University  of California Observatories. KAIT and its ongoing operation were made possible by donations from Sun Microsystems, Inc., the   Hewlett-Packard Company, AutoScope Corporation, Lick Observatory, the NSF, the University of California, the Sylvia \& Jim Katzman Foundation, and the TABASGO Foundation. We thank Kate Bostow, Asia DeGraw, Gracelynn Jost, Audrey Liddle, Andrew McHaty, Katherine Mora, Arnav Swaroop, William Wu, and Emma Yu for their contributions to the optical photometric observations with the Nickel telescope. Research at Lick Observatory is partially supported by a gift from Google.

We acknowledge the use of the large language models ChatGPT \citep{openai_chatgpt_2026} and Claude \citep{anthropic_claude_sonnet5_2026, anthropic_claude_opus5_2026} for assistance with editing and revising, and code development.

\end{acknowledgements}

\begin{contribution}

\textbf{Conceptualization:} L.A.K., S.W.J., S.B., J.E.A., K.A.B., A.V.F., A.F., R.J.F., O.G., C.K., K. Maeda, J.P., H.S., T.S., T.T., S.V.

\textbf{Analysis:} L.A.K., S.B., A.A.M., S.W.J., C. Larison, J.V.

\textbf{Data:} L.A.K., S.W.J., W.B.H., C.M.P., E.Z.A., J.E.A., M.A., C.A., A.B., K.A.B., T.G.B., C.T.C., J.R.F., A.V.F., E.G., D.A.H., D.O.J., R.K., M.K., C. Liu, K. Maeda, C.M., K. Medler, N.E.M., A.M., A.P., R. Patlak, J.P., A.P.R., N.R., D.J.S., B.J.S., M. Shrestha, K.T., J.H.T., S.V., K.W., Y.Y., W.Z.

\textbf{Software:} L.A.K., S.B., M.A., K.A.B., J.R.F., C.M., S.V., K.W.

\textbf{Visualization -- review \& editing:} L.A.K., S.B., A.A.M., S.W.J., C.Liu, K. Medler.

\textbf{Writing -- review \& editing:} L.A.K., S.B., A.A.M., S.W.J., W.B.H., K.A., A.V.F., A.F., J.T.H., C.K., C. Larison, C. Liu, K. Maeda, K. Maguire, R. Pakmor, N.R., A.R., D.J.S., M. Shrestha, M. Singh, T.S., J.C.W., Y.Y.

\textbf{Interpretation:} L.A.K., S.B., A.A.M., S.W.J., W.B.H., F.P.C., J.T.H., C. Liu, K. Maguire, R. Pakmor, D.J.S.

\end{contribution}

%
\facilities{JWST (NIRSpec and MIRI), IRTF (SpeX), Gemini (GMOS, GNIRS), NOT (ALFOSC), Lick (Shane/Kast, Nickel, KAIT), MMT (Binospec), Magellan Baade (FIRE), LCOGT}

\software{Astropy \citep{astropycollaboration_astropy:_2013, astropycollaboration_astropy_2018, AstropyCollaboration2022}, 
Matplotlib \citep{hunter_matplotlib:_2007}, 
NumPy \citep{oliphant_guide_2006}, dust extinction \citep{karl_gordon_2022_6397654}, jwst \citep{Bushouse_JWST_Calibration_Pipeline_2022}, CMFGEN \citep{Hillier2012}, ChatGPT \citep{openai_chatgpt_2026}, Claude \citep{anthropic_claude_sonnet5_2026, anthropic_claude_opus5_2026}
}


\appendix

\section{MIR Line List \label{sec:appendix}}

\autoref{tab:lineIDs_25rbs} lists the MIR permitted and forbidden transitions marked in \autoref{fig:jwst_spec}. This list is not exhaustive; other weaker or unidentified lines may contribute. The full CMFGEN model line lists corresponding to the models shown in \autoref{fig:models} and \autoref{fig:Mg_merger} are available upon request.

\begin{table}
\centering
\caption{Line identifications marked in \autoref{fig:jwst_spec} for \rbs\ between 3 and 14\um.}
\label{tab:lineIDs_25rbs}
\begin{tabular}{lr@{\hspace{1.5em}}|@{\hspace{1.5em}}lr}
\hline\hline
Line & $\lambda$ ($\mu$m) & Line & $\lambda$ ($\mu$m) \\
\hline
{[}\ion{Fe}{3}{]} & 3.043 & \ion{Si}{2} & 5.226 \\
\ion{Mg}{2}$^{a}$ & 3.090 & \ion{Si}{2} & 5.272 \\
\ion{Si}{3} & 3.139 & \ion{Si}{2} & 5.396 \\
{[}\ion{Ca}{4}{]} & 3.206 & {[}\ion{Co}{4}{]} & 5.416 \\
{[}\ion{Ni}{3}{]} & 3.393 & {[}\ion{Ni}{2}{]} & 6.634 \\
{[}\ion{Co}{3}{]} & 3.491 & {[}\ion{Ar}{2}{]} & 6.983 \\
\ion{Si}{2} & 3.494 & {[}\ion{Ni}{3}{]} & 7.347 \\
\ion{Ni}{2} & 3.541 & {[}\ion{Fe}{3}{]} & 7.788 \\
{[}\ion{Ni}{3}{]} & 3.801 & {[}\ion{Ni}{4}{]} & 8.405 \\
\ion{Si}{3} & 3.887 & {[}\ion{Ar}{3}{]} & 8.989 \\
\ion{Si}{2} & 3.947 & \ion{Mg}{2}$^{a}$ & 9.708 \\
\ion{Mg}{2}$^{a}$ & 4.049 & {[}\ion{S}{4}{]} & 10.508 \\
\ion{Ni}{2} & 4.079 & {[}\ion{Co}{2}{]} & 10.520 \\
\ion{Mg}{2}$^{a}$ & 4.109 & {[}\ion{Ni}{2}{]} & 10.679 \\
\ion{Ni}{2} & 4.310 & {[}\ion{Ni}{3}{]} & 10.999 \\
\ion{Mg}{2}$^{a}$ & 4.539 & {[}\ion{Co}{3}{]} & 11.885 \\
\ion{Mg}{2}$^{a}$ & 4.761 & \ion{Mg}{2}$^{a}$ & 13.120 \\
\hline
\end{tabular}
\begin{minipage}{0.8\linewidth}
\vspace{0.5ex}
\footnotesize
\textbf{Note.} Wavelengths are vacuum values. Lines in square brackets are forbidden transitions, those without are permitted transitions. \\
$^{a}$ \ion{Mg}{2} lines are permitted-like transitions whose upper levels are combinations of multiple high-$\ell$ states (see \autoref{sec:mir_evolution}).
\end{minipage}
\end{table}




\bibliography{references}{}
\bibliographystyle{aasjournalv7}



\end{document}